\documentclass[draft]{agujournal}

\journalname{Journal of Geophysical Research}

\begin{document}

%
%

\title{Sustenance of phytoplankton in the subpolar North Atlantic during the winter through patchiness}

%
%

 \authors{Farid Karimpour\affil{1},
 Amit Tandon\affil{1}, and Amala Mahadevan\affil{2}}

\affiliation{1}{Department of Mechanical Engineering,
University of Massachusetts, Dartmouth, Massachusetts 02747, USA.}
\affiliation{2}{Woods Hole Oceanographic Institution, Woods Hole, Massachusetts 02543, USA.}


\correspondingauthor{Farid Karimpour}{fkarimpour@umassd.edu}


\begin{keypoints}
\item In the subpolar North Atlantic even during the winter the fronts restratify the mixed layer, in spite of strong winds and loss of heat.
\item The phytoplankton growth considerably increases in the presence of fronts compared to cases without fronts.
\item Highly variable air-sea fluxes have little effect on the phytoplankton growth in the subpolar North Atlantic during the winter.
\end{keypoints}

%
%

\begin{abstract}
This study investigates the influence of two factors that change the mixed layer depth and can potentially contribute to the phytoplankton sustenance over winter: 1) variability of air-sea fluxes and 2) three-dimensional processes arising from strong fronts.   To study the role of these factors, we perform several three-dimensional numerical simulations forced with air-sea fluxes at different temporal averaging frequencies as well as different spatial resolutions. Results show that in the winter, when the average mixed layer is much deeper than the euphotic layer and the days are short, phytoplankton production is relatively insensitive to the high-frequency variability in air-sea fluxes. The duration of upper ocean stratification due to high-frequency variability in air-sea fluxes is short and hence has a small impact on phytoplankton production. On the other hand, slumping of fronts creates patchy, stratified, shallow regions that persist considerably longer than stratification caused by changes in air-sea fluxes. Simulations show that before spring warming, the average MLD with fronts is about 700 m shallower than the average MLD without fronts. Therefore, fronts increase the residence time of phytoplankton in the euphotic layer and contribute to phytoplankton growth. Results show that before the spring warming, the depth-integrated phytoplankton concentration is about twice as large as phytoplankton concentration when there are no fronts. Hence, fronts are important for setting the MLD and sustaining phytoplankton in the winter. Model results also show that higher numerical resolution leads to stronger restratification, shallower mixed layers, greater variability in the MLD and higher production of phytoplankton.
\end{abstract}


%
%

%



%
%
%

\section{Introduction}
\label{sec:intro}
Phytoplankton are one of the lowest end members of the food web in the ocean and hence have a first-order effect on sustaining life in aquatic environments. 
Like plants, phytoplankton life relies on photosynthesis, while unlike plants that are fixed, they are adrift and are substantially affected by the ocean currents (see \citet{MAHADEVAN[2016]} for a recent review).
Phytoplankton use solar radiation to consume carbon dioxide and nutrients during photosynthesis and as a result, generate oxygen. The consumption of ocean carbon dioxide by phytoplankton allows more carbon dioxide, a key greenhouse gas, to be dissolved in the ocean from the atmosphere. 
Therefore, phytoplankton have a substantial impact on the ecosystem by  photosynthetic fixation of the carbon and contributing to the ocean uptake of the rapidly accumulated $\rm{CO_2}$ in the atmosphere.
The North Atlantic, well-known for its springtime phytoplankton bloom, is one of the biologically active regions of the ocean. It substantially contributes to the photosynthetic fixation of the carbon and it is estimated that the North Atlantic is responsible for about 25\% of the net $\rm{CO_2}$ uptake from the atmosphere. (\citet{SABINE[2004], TAKAHASHI[2009]}).

There are different mechanisms and parameters, which control the production of  phytoplankton (\citet{MILLER[2012]}), however, availability of light and nutrients are considered to be the major drivers for primary production. During the winter, nutrients are replete in the ocean (\citet{TOWNSEND[1994]}) and hence phytoplankton production is not limited by nutrient availability (\citet{SVERDRUP[1953]}), but rather by the low-level sunlight of short winter days. 
Strong cooling and wind forcing cause the upper ocean surface to actively mix to a few hundred meters, creating a deep oceanic mixed layer (ML). The induced turbulent, convective  motions deprive phytoplankton of the limited sunlight and  photosynthetically available radiation (PAR) at the surface during the winter in the subpolar North Atlantic. This considerably inhibits the growth of phytoplankton, leading to decrease in the phytoplankton population during the winter. It is the sustenance of the phytoplankton population in the subpolar North Atlantic during the winter which sets the initial conditions for the rapid, exponential growth and bloom of phytoplankton in the early spring. Therefore, it is very important to know what physical mechanisms help sustain the phytoplankton population through the winter.

Different scenarios contributing to the sustenance of phytoplankton have been proposed. \citet{RILEY[1949]} studied the sustenance of phytoplankton by deriving a relationship between the turbulence and sinking velocity without considering the light dependence of phytoplankton. \citet{SVERDRUP[1953]} studied the role of light on phytoplankton growth and proposed his well-known critical depth hypothesis. This study argues that phytoplankton can grow when the mixed layer depth (hereafter MLD) is shallower than a critical depth. Sverdrup's model assumes that the mixed layer is well mixed and nutrients are abundant. The suitability of his proposition has become a matter of debate as other researchers such as \citet{BEHRENFELD[2010]} and \citet{BOSS[2010]} observed growth in deep winter mixed layers deeper than the critical depth.  \citet{BEHRENFELD[2010]}  argued that phytoplankton concentration is diluted when the mixed layer deepens resulting in decrease of encounter between phytoplankton and grazers. In this case, maximum growth rate is seen when MLD is deep.  
\citet{HUISMAN[1999], HUISMAN[2002]} argued that phytoplankton can survive when there is an intermediate level of turbulence in the water column. They introduced a  maximal (critical) and minimal turbulence level for the production of phytoplankton. When the level of turbulence, which is measured in terms of turbulent diffusivity, is less than the critical turbulence level, phytoplankton growth dominates the turbulent mixing and they can survive. On the other hand, the level of turbulence should be strong enough to overcome the sinking velocity, keep phytoplankton in the euphotic layer and prevent  them from sinking to the dark, deep ocean. Later, \citet{BACKHAUS[2003]} and \citet{DASARO[2008]} discussed that the scenario presented by \cite{HUISMAN[1999], HUISMAN[2002]} is incomplete as the convection caused by cooling of the ocean surface during the winter is not incorporated. They argued that convection forms cells with orbital motions in the so-called convective mixed layer (CML). These orbital motions enable phytoplankton in the CML to be transported from deep layers of the ocean to the euphotic layer (and vice versa) and hence can  potentially sustain growth. This implies that convection creates a virtual euphotic layer in which the phytoplankton production is sustained in spite of being deeper than the actual euphotic layer. More recently, \citet{TAYLOR[2011b]} discussed that the shutdown of convection reduces turbulent mixing and allows phytoplankton to grow before  the spring warming.  
\citet{BRODY[2014]} evaluated the bloom by developing a mixing length scale based on buoyancy forcing and wind stress. They concluded that the bloom begins when the mixing length scale is shallower than euphotic layer. They challenged the critical turbulence hypothesis by considering mixing due to convection and wind and discussed bloom initiates when the mixing is due to wind rather than convection.

The hypotheses mentioned above provide reasonable explanations of how phytoplankton persist in the winter through one-dimensional processes, however without a detailed investigation of how air-sea fluxes affect phytoplankton production. In addition, fronts are ubiquitous in the real ocean and are sources of several forms of instabilities and three-dimensional processes, which can result in increased vertical stratification. Also, while winter conditions are generally harsh in the subpolar ocean,  periods of reduced wind stress  and increased heat flux have been observed in the North Atlantic (\citet{LACOUR[2017]}). In a recent study, \citet{LACOUR[2017]} investigated blooms during the winter in the subpolar North Atlantic and found  that regions of shallow MLD (i.e. MLD<100 m),  formed by mixed layer eddies arising from fronts, are very common in the ocean during the winter. Their observations showed that patches of different stratification have distinct light environments, where light level in the stratified shallow patches is considerably higher than deep regions leading to growth of phytoplankton in about 70\% of stratified regions.  

The mixed layer is sustained through a competition between processes that increase the turbulent mixing and processes that restratify the mixed layer (\citet{MAHADEVAN[2010]}). Air-sea fluxes have a first-order effect on the turbulence and mixed layer. While  convective fluxes due to cooling, evaporation or down-front winds, which push heavier water over lighter water, increase turbulent mixing, destroy the stratification and deepen the mixed layer,  heating, fresh water and up-front winds can regenerate the stratification of the mixed layer. The modulation of turbulent mixing and stratification by air-sea fluxes will eventually affect phytoplankton production. 
 Strong winter storms or cooling events destroy the stratification, deepen the ML and dilute phytoplankton concentration. On the contrary, physical processes that increase the stratification or shoal the MLD such as warming events and up-front winds, increase the residence time of phytoplankton in the euphotic layer. The role of these intermittent air-sea fluxes on the sustenance of phytoplankton during the winter will be discussed in this article.
 



Mesoscale eddies which are geostrophic and primarily two-dimensional contribute to the evolution and production of phytoplankton (\citet{MCGILLICUDDY[2016]}). In addition, recent studies have shown a chain of three-dimensional processes associated with frontal instabilities occurs in the ocean with  length scales $\mathcal{O}$(0.1-10 km) and time scales of a few days (\citet{BOCCALETTI[2007]}, \citet{FOXKEMPER[2008]}, \citet{MAHADEVAN[2012]}). Phytoplankton growth time scales are on the order of days and hence can get influenced by frontal instabilities (\citet{MAHADEVAN[2012]}). Fronts have been recognized to be hotspots for enhancement of phytoplankton concentration (\citet{FRANKS[1992]}). Frontal processes leading to growth of phytoplankton have been investigated in recent studies, for example by \citet{TAYLOR[2011b]}, \citet{MAHADEVAN[2012]} and \citet{WHITT[2017]}.  
Frontal instabilities and the resultant vertical restratification  cannot be captured in conventional models, which merely rely on one-dimensional budgets for prediction of the mixed layer depth (MLD). The mixed layer is maintained through surface fluxes, vertical mixing, and both lateral and vertical processes, which need to be investigated through three-dimensional numerical simulations that are able to properly resolve submesoscale processes.


In this study, we aim to answer three main questions regarding phytoplankton production and sustenance in the North Atlantic during the winter in the presence of  fronts and forced by realistic high-frequency air-sea fluxes:
1) Do episodic, high frequency air-sea fluxes influence phytoplankton productivity?
2) How do fronts contribute to sustenance of phytoplankton in the winter? This is an important question considering the proven role of fronts in stratification and since winter is an active season for submesoscale processes (\citet{CALLIES[2015]}, \citet{LUO[2016]}).
3) What is the effect of frontal strength, numerical resolution and inclusion of submsesocale fronts on the mixed layer depth, phytoplankton and their non-uniform distribution in the ocean? 
In order to answer these questions we simulate the dynamics of the subpolar North Atlantic upper ocean in the winter, which is under the influence of highly variable air-sea fluxes and stratifying submesoscale eddies arising from fronts. To this end, we perform highly resolved three-dimensional numerical simulations that can capture frontal instabilities using the Process Study Ocean Model (PSOM, \citet{MAHADEVAN[2006]}) forced by high frequency (hourly), strong winter air-sea fluxes. We evaluate upper ocean physical-biological interactions  and address the questions related to sustenance of  phytoplankton in the turbulent ocean during the winter. 

The outline of this paper is as follows. In section \ref{sec:num_sim} we discuss the problem set-up and the numerical approach. In section \ref{sec:results} we discuss the results from numerical simulations and show that in spite of the fact that growth conditions are unfavorable during the winter, phytoplankton can survive mainly due to restratification and growth associated with frontal instabilities. Section \ref{sec:conc} presents the conclusions for this study. 

\section{Numerical simulations}
\label{sec:num_sim}
In this section we  discuss the numerical code, the model setup, initial and boundary conditions, and  the turbulence closure scheme.

\subsection{Numerical model and turbulence closure scheme}
\label{sec:code}
We use the Process Study Ocean Model (PSOM, \citet{MAHADEVAN[2006]}), a three-dimensional model that solves the Reynolds averaged Navier-Stokes (RANS) momentum and tracer equations.
The Reynolds decomposition and averaging of the fluid flow governing equations lead to extra sub-grid scale turbulent momentum fluxes ($\overline{u'_i u'_j}$) and turbulent scalar fluxes ($\overline{u'_i c'}$), which cannot be solved directly and are parameterized in the numerical model (\cite{POPE[2000]}). 
Here, $i$ and $j=1,2,3$ represent $x$, $y$ and $z$ directions, $\overline{(\hspace{1.5 mm})}$ denotes the spatial or temporal average and $u'$ and $c'$ are turbulent velocity and turbulent scalar fields, respectively. 
For sub-grid scale fluxes, PSOM is coupled with a suite of turbulence models called the General Ocean Turbulence Model (GOTM) developed by \citet{BURCHARD[1999]}. GOTM gives the opportunity to benefit from more sophisticated closure schemes, such as the $k$-$\epsilon$ turbulence model (\citet{JONES[1972]}) or K-Profile Parameterization (KPP) model (\citet{LARGE[1994]}) in our simulations.

PSOM is coupled with GOTM such that it solves the discretized Reynolds-averaged Navier-Stokes (RANS) equations while relying on GOTM for the calculation of the turbulent fluxes. In our simulations, we have used the standard $k$-$\epsilon$ closure scheme  (\citet{JONES[1972]}).
 For calculation of turbulent momentum fluxes, PSOM uses the turbulent viscosity hypothesis which links the turbulent momentum fluxes to the mean gradient of the velocity as 
\begin{equation}\label{eq:nu_t}
\overline{u_i'u'_j} = -K_{ij}^m \overline{S}_{ij},
\end{equation}
where $K_{ij}^m$ is the turbulent (eddy) viscosity and $\overline{S}_{ij}= 1/2 \left({ \partial {\overline{U}_i}/ \partial x_j + \partial{\overline{U}_j} /\partial x_i} \right)$ is the mean strain rate. The turbulent viscosity is calculated by GOTM and passed to PSOM. 
In the $k$-$\epsilon$ closure scheme the vertical turbulent viscosity ($K^m_z$) is calculated as a function of the turbulent kinetic energy $(k)$ and the dissipation rate of the turbulent kinetic energy $(\epsilon)$ as  
\begin{equation}\label{eq:Kz_keps}
 K^m_z= C_\mu \frac{k^2}{\epsilon},
\end{equation}
where $C_\mu=0.09$ is the intensity ratio constant. 
In this model, $k$ and $\epsilon$ are obtained by solving separate evolution equations for each of them. These equations incorporate advection terms, the effect of production and destruction of turbulence, buoyancy forces and inhomogeneity arising from the wall (\citet{KVJGR[2014]}).  The $k$-$\epsilon$ model has been tested extensively and is considered the most widely used turbulence model (\citet{POPE[2000]}). In our study, it is vital to properly incorporate the effects of highly variable air-sea fluxes and buoyancy forces on the turbulent mixing. Considering the level of completeness of the $k$-$\epsilon$ closure scheme, it is a suitable model for calculating turbulent viscosity in our simulations.

To calculate the mixing of the tracer (i.e. density, heat, salinity, etc.), PSOM uses the gradient-diffusion hypothesis, which assumes that the turbulent scalar fluxes are transported down the mean scalar gradient and are calculated as
\begin{equation}\label{eq:kappa_t}
\overline{u_j'c'} = -K_{j}^d \frac{\partial \overline{C}}{\partial x_j},
\end{equation}
where $K_j^d$ is the turbulent diffusivity. 

In our simulations, we assume that the horizontal turbulent viscosity and diffusivity are isotropic (i.e. invariant under rotation), equal and constant as $K_x^m=K_y^m=K_x^d=K_y^d=0.1~m^2/s$. However, this is not the case for the vertical components. The vertical turbulent viscosity ($K^m_z$) is calculated through equation \ref{eq:Kz_keps} and the turbulent diffusivity is calculated as $K_z^d=K_z^m/Pr_t$, where $Pr_t$ is  the turbulent Prandtl number, which is the linking bridge between the turbulent viscosity and diffusivity. There are different propositions based on the mean flow quantities such as the mean shear rate and buoyancy frequency for the turbulent Prandtl number. We use the $Pr_t$ parameterization proposed by \citet{VS[2010]}, which is a slightly modified formulation of \citet{SCHUMANN[1995]}. Its suitability has been  verified against Direct Numerical Simulation (DNS) data and is formulated as follows
\begin{equation}\label{eq:Prt_vs}
Pr_t = \frac{Ri_g}{R_{f \infty}} + Pr_{t0} \left (-\frac {Ri_g}{Pr_{t0} \Gamma_{\infty}} \right).
\end{equation}
Here, $Ri_g = N^2/S^2$ is the gradient Richardson number, $N$ is the buoyancy frequency and $S$ is the horizontal mean velocity gradient (i.e. mean shear rate) in the vertical direction. $R_f$ is the flux Richardson number and $R_{f \infty} = 0.25$ and  $\Gamma_{\infty} = 1/3$ are the flux Richardson number and the mixing efficiency when $Ri_g \rightarrow \infty$. Also, $Pr_{t0}=0.7$ is the neutral turbulent Prandtl number, where there is no density stratification. The dynamic nature of this proposition allows for the constant adjustment of the turbulent diffusivity according to the dynamical and buoyancy forces in the flow and allows for a more realistic modeling of the mixing of the relevant scalar field.

\begin{figure}
	\centering
	{\includegraphics[width=6.in]{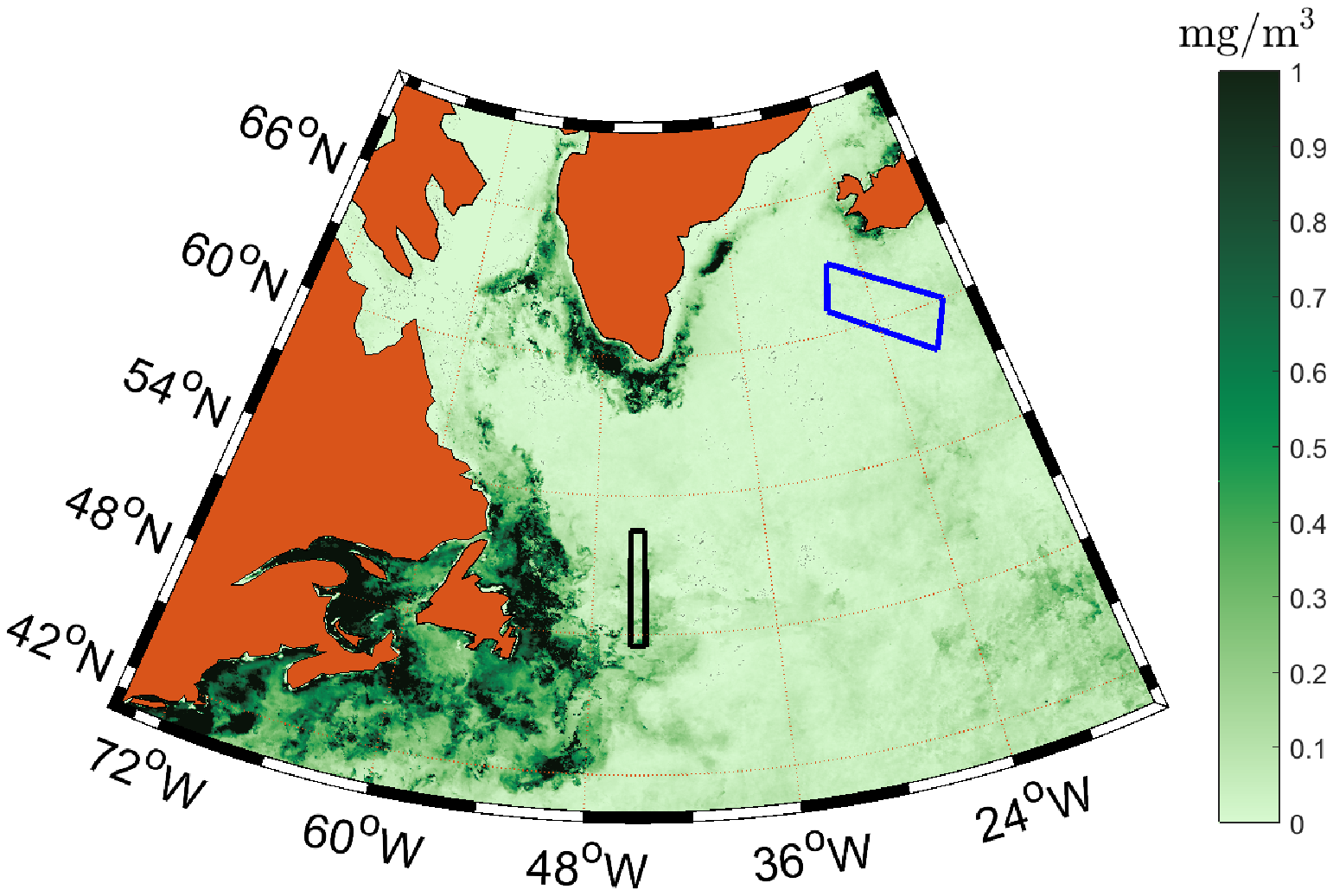}}
	\put(-365,240){(a)} \\
		{\includegraphics[width=4in]{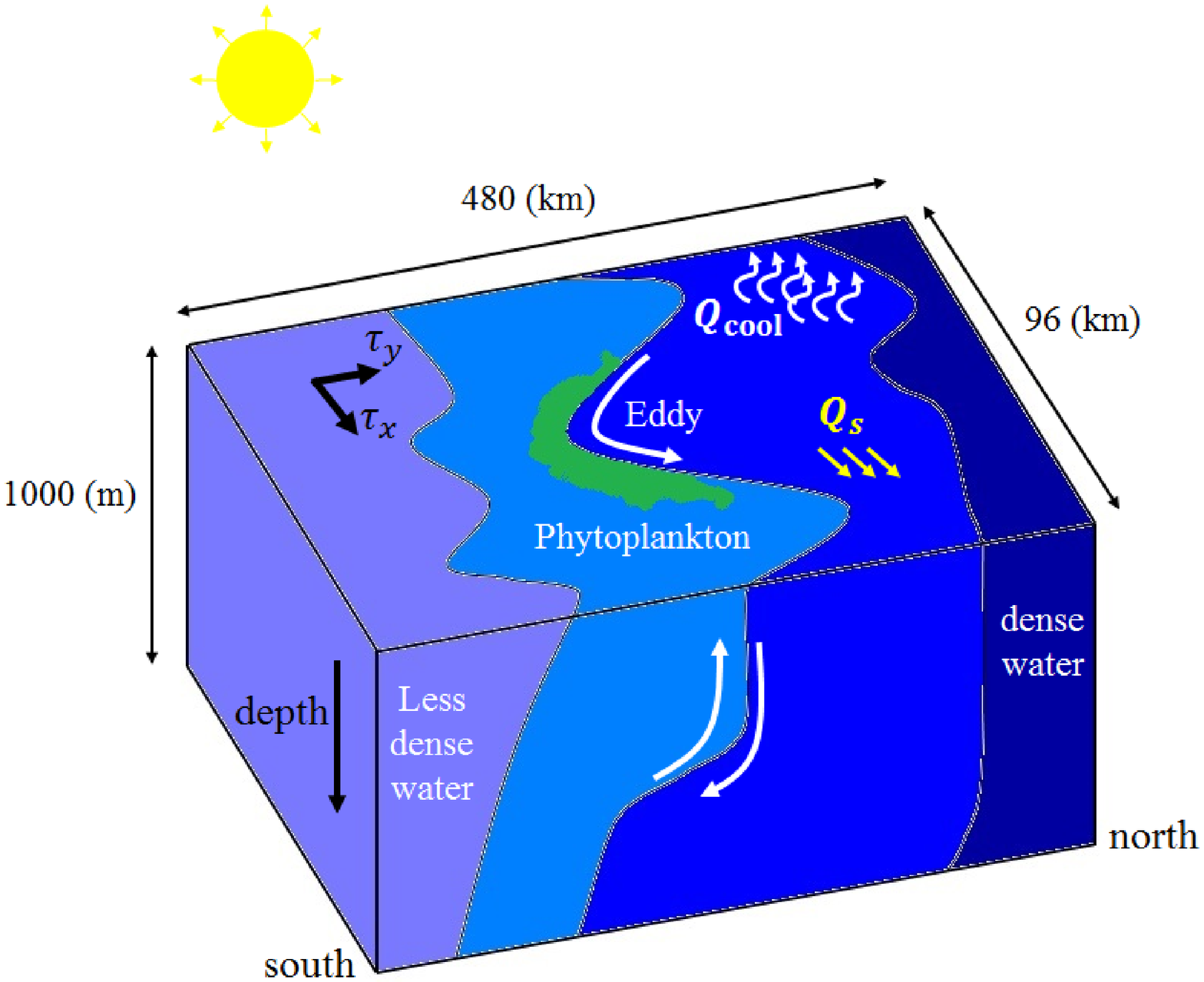}}
		\put(-335,200){(b)} \\ 
	\caption{(a) The approximate location of the modeled region in the North Atlantic shown by the black, solid line box. The green color shows the  concentration of Chlorophyll-a obtained from MODIS-Aqua averaged from yeardays 50 -- 150. Also, the North Atlantic Bloom Experiment 2008 (\citet{FENNEL[2011]}) approximate location (solid, blue box) which was done in April-June 2008 is shown for comparison. (b) The schematic of the modeled domain in the North Atlantic.}
	\label{fig:schematic}
\end{figure}

\subsection {Domain description} \label{sec:domain}

We use a three-dimensional model domain for our simulations in PSOM which is $96~\rm{km}~\times~480~\rm{km}~\times~1000~\rm{m}$, in $x$ (or zonal), $y$ (or meridional) and $z$ (or vertical) directions respectively, with $n_x=96$, $n_y=480$ and $n_z=48$ grid cells in the zonal, meridional and vertical directions.
We use a horizontal resolution of $\Delta x=\Delta y=1000 ~\rm{m}$ to allow resolving submesoscale processes and a stretched grid in the vertical direction with resolution ranging from about 4.6 m at the surface to 47 m at the bottom of the domain.
The zonal direction uses periodic boundary conditions while in the meridional direction the boundaries are solid walls. The model uses a time step of 432 s to evolve the momentum and scalar evolution equations. An approximate location of the modeled domain in the North Atlantic is shown in Figure \ref{fig:schematic}a. The blue box on the right corner of Figure \ref{fig:schematic}a shows the location of the North Atlantic Bloom Experiment, which was conducted in April-June 2008 to study the spring bloom to the south of Iceland (\citet{FENNEL[2011]}). Figure  \ref{fig:schematic}b shows the schematic of the modeled domain.

\subsection {Atmospheric forcing} \label{sec:atm_force}
\begin{figure}
	\centering
	{\includegraphics[width=5.1in]{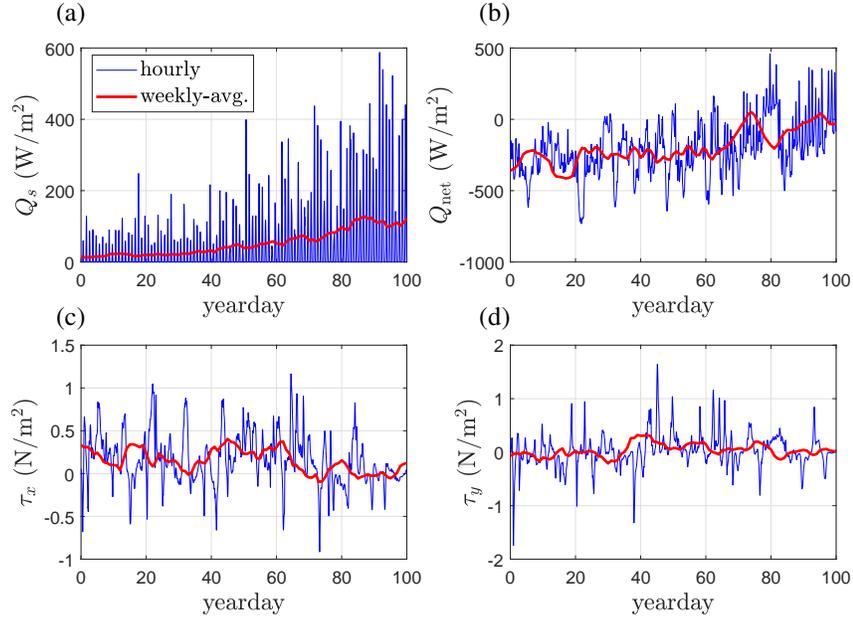}}
 	\put(-330,230){(a)} \put(-170,230){(b)} \put(-330,115){(c)}  \put(-170,115){(d)}\\ 
	\caption{The hourly and weekly-averaged fluxes for (a) the shortwave radiation ($Q_s$);
	(b) the net heat flux ($Q_{\rm{net}}$); 
	(c) the zonal wind stress ($\tau_x$); 
	(d) the meridional wind stress ($\tau_y$); for the region shown in Figure \ref{fig:schematic}.}
	\label{fig:forcing}
\end{figure}

The numerical model is forced by NASA's Modern-Era Retrospective analysis for Research and Applications, Version 2 (MERRA-2) data for 2008. MERRA-2 data are satellite-based reanalysis data which are continuous in time and space and are produced with the  Global Modeling Assimilation Model Office/Goddard Earth Observing System Model, Version 5 (GMAO/GEOS-5) data assimilation system (\citet{BOSILOVICH[2015]}). MERRA-2 reanalysis that we use are hourly and are averaged in the range  $51^{\circ} \rm{N}- 53^{\circ} \rm{N}$ with a spatial resolution of $0.5^{\circ}$ and $44.375^{\circ} \rm{W}- 45.625^{\circ} \rm{W}$, with a spatial resolution of $0.625^{\circ}$, corresponding to the region shown in Figure \ref{fig:schematic}a with the black box. The data include the wind stress in both the zonal and meridional directions, the shortwave radiation ($Q_s$), longwave radiation ($Q_{\rm{long}}$), the sensible ($Q_{\rm{sens}}$) and latent ($Q_{\rm{lat}}$) heats. Figure \ref{fig:forcing} shows the hourly fluxes obtained from MERRA-2 for the time ranging from yearday 0 to 100 of 2008 and includes the shortwave (\ref{fig:forcing}a), the net heat flux (\ref{fig:forcing}b) which is the summation of the shortwave and longwave radiative, sensible and the latent heat fluxes (i.e.  $Q_{\rm{net}}= Q_{s} + Q_{\rm{long}} + Q_{\rm{sens}} + Q_{\rm{lat}}$). The zonal wind stress ($\tau_x$) and meridional wind stress ($\tau_y$) are shown in Figures \ref{fig:forcing}c and d. In Figure \ref{fig:forcing}, the weekly-averaged data is superimposed on the hourly data. In our simulations, we impose a linear gradient in $Q_{\rm{net}}$ such that there is a 50 $\rm{W/m^2}$ difference between the southern and northern boundaries. At the southern boundary, the imposed $Q_{\rm{net}}$ is +25 $\rm{W/m^2}$ more than the values shown in Figure \ref{fig:forcing}b and at the northern boundary, the imposed $Q_{\rm{net}}$ is -25 $\rm{W/m^2}$ less than the values shown in Figure \ref{fig:forcing}b.
%
The average and standard deviation of hourly $Q_s$, the hourly and weekly-averaged $Q_{\rm{net}}$, $\tau_x$ and $\tau_y$ are presented in Tables \ref{tab:forcing_cool} and  \ref{tab:forcing_wind}, which show a high standard deviation for hourly fluxes due to their high variations. 

Having a closer look at both the hourly and  averaged data in Figure \ref{fig:forcing} shows that hourly $Q_s$, $Q_{\rm{net}}$ and wind stress can be much greater or smaller than averaged data. Such conditions can cause distinct behaviors in both momentum and scalar mixing, restratification and potentially phytoplankton production which cannot be observed when using smoothed fluxes. 

\begin{table}
	\caption{Mean and standard deviation of the hourly shortwave radiation ($Q^{\rm{hourly}}_s$), net hourly heat flux ($Q^{\rm{hourly}}_{\rm{net}}$) and net weekly-averaged heat flux ($Q^{\rm{weekly}}_{\rm{net}}$) near the studied region for the first 100 days of 2008.}
	\centering
	\begin{tabular}{l c c}
		\hline
		Flux  & Average $\rm {(W/m^2)}$ & std $\rm {(W/m^2)}$ \\
		\hline
		$Q_s^{\rm{hourly}}$  & 47.97  & 90.98 \\
		$Q_{\rm{net}}^{\rm{hourly}}$  & -212.31  & 189.98 \\
		$Q_{\rm{net}}^{\rm{weekly}}$  & -202.72  & 114.68 \\
		\hline
	\end{tabular}
	\label{tab:forcing_cool}
\end{table}

\begin{table}
	\caption{Mean and standard deviation of the hourly zonal wind stress ($\tau^{\rm{hourly}}_x$), the hourly meridional wind stress ($\tau^{\rm{hourly}}_y$), the weekly-averaged zonal wind stress ($\tau^{\rm{weekly}}_x$), the weekly-averaged meridional wind stress ($\tau^{\rm{weekly}}_y$) near the studied region for the first 100 days of 2008.}
	\centering
		\begin{tabular}{l c c}
			\hline
			Flux  & Average $\rm {(N/m^2)}$ & std $\rm {(N/m^2)}$ \\
			\hline
			$\tau_x^{\rm{hourly}}$  & 0.158  & 0.297 \\
			$\tau_x^{\rm{weekly}}$  & 0.153  & 0.137 \\
			$\tau_y^{\rm{hourly}}$  & 0.025  & 0.30 \\
			$\tau_y^{\rm{weekly}}$  & 0.032  & 0.12 \\
			\hline
		\end{tabular}
		\label{tab:forcing_wind}
	\end{table}

\subsection {Initial density and fronts} \label{sec:init_dens}

\begin{figure}
	\centering
	{\includegraphics[width=2.6in]{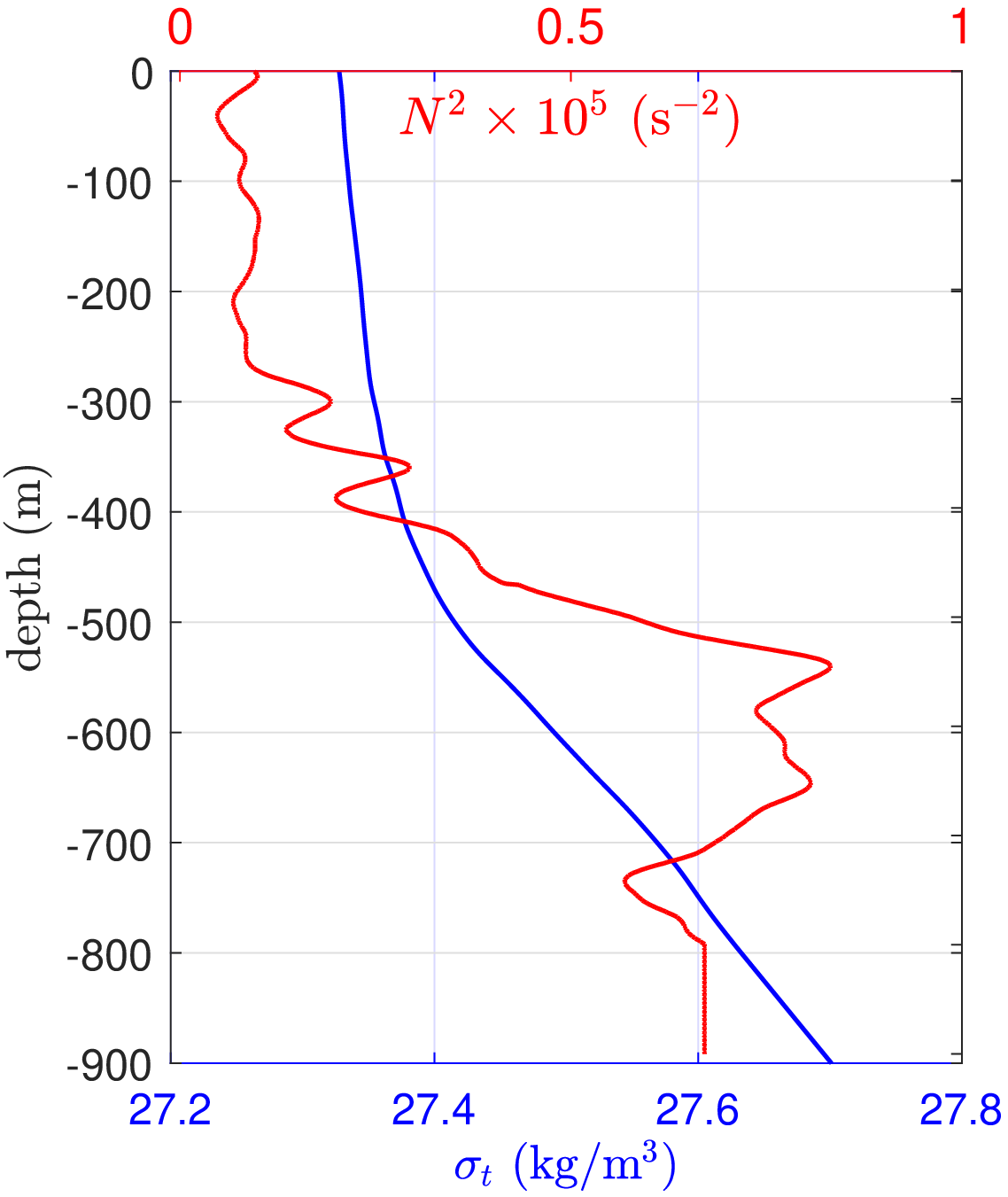}}
	{\includegraphics[width=2.6in]{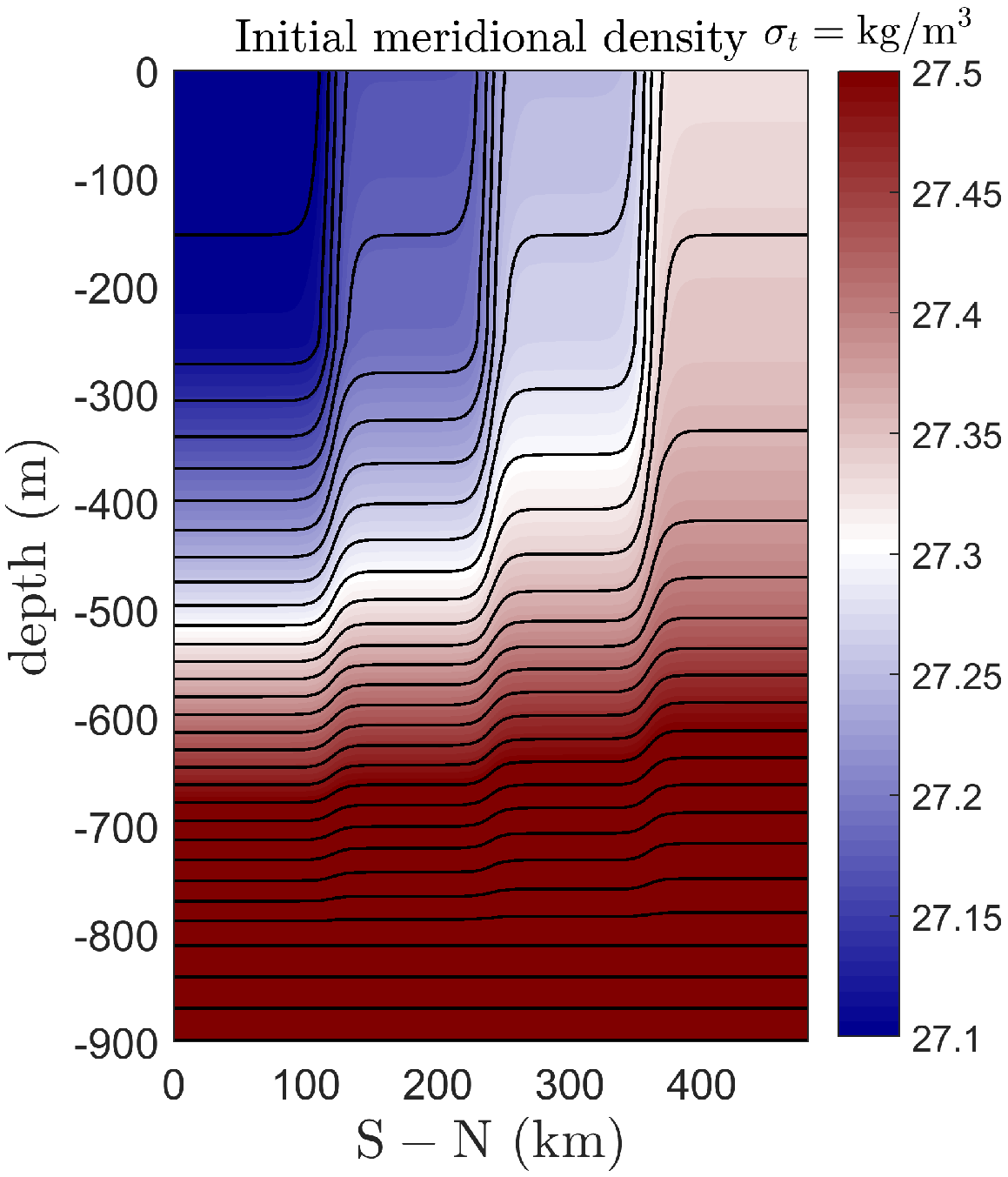}}
	\put(-365,210){(a)} \put(-175,210){(b)} \\

	\caption{(a) The initial density profile and the initial buoyancy frequency. Below $z\approx-800~\rm{m}$ the density is linear and hence $N^2$ is constant; (b) The initial density distribution in the meridional ($y$) direction, the solid black lines show the contours of density. Here, $\sigma_t=(\rho-1000)~\rm{(kg/m^3)}$ is the density anomaly.}
	\label{fig:init_den}
\end{figure}

The initial density profile data is the same as the density profile used in the simulations of \citet{MAHADEVAN[2012]}  as shown in Figure \ref{fig:init_den}a. Here for the sake of simplicity, the density is represented by a density anomaly $\sigma_t = \rho - \rho_0$ where $\rho$ is the potential density and $\rho_0=1000~\rm{kg/m^3}$. The potential density profile was obtained from Argo floats that collected data in south of Iceland between 2000 and 2009. Figure \ref{fig:init_den}b shows the initial buoyancy frequency which is a measure of the density stratification strength and is calculated as
\begin{equation}\label{eq:N2}
N^2=\frac{\partial b}{\partial z} =\frac{-g}{\rho_0}  \left(\frac{\partial \sigma_t}{\partial z}\right),\,
\end{equation} 
 where $b = (-g/\rho_0) (\rho- {\rho_0}$) is the buoyancy. $N^2$ is constant below  $z\approx 800~\rm{m}$. 

Also,  Figure \ref{fig:init_den}b illustrates the distribution of the density anomaly  in the meridional direction and shows the density increases towards the north. 
Three fronts  span the domain in the meridional direction. 
The maximum horizontal buoyancy gradient $M^2_y =(-g/\rho_0)(\partial  {\sigma_t}/ \partial y) \approx 3.6 \times 10^{-8}~\rm{s^{-2}}$ which is close to the density gradient observed in North Atlantic Bloom Experiment (NAB08) conducted in 2008 (\citet{FENNEL[2011]}, \citet{MAHADEVAN[2012]}).

\subsection{Biological model} \label{sec:bio_model}
A phytoplankton evolution model proposed by \citet{BAGNIEWSKI[2011]} is coupled to PSOM  and is described by
\begin{equation}\label{eq:phy_eq}
\frac{\partial P}{\partial t} + \nabla. ({\bf u} P) = \mu_{P} P - m_P P  + \frac{\partial }{\partial z}\left(w_s P \right) + \frac{\partial }{\partial z} \left( K_z^d \frac{\partial P}{\partial z}\right). 
\end{equation}
In this equation, $P$ is the phytoplankton concentration and ${\bf u}= (u,v,w)$ is the velocity vector. $\mu_{P}$ is the growth rate, $m_{P}=0.0187~\rm{d^{-1}}$ is the mortality rate and $w_s=1.2~\rm{m/day}$ is the sinking rate of phytoplankton. 

The growth rate is calculated as 
\begin{equation} \label{eq:growth_rate}
\mu_P = \mu_P^{\rm{max}} \frac{I_P \alpha_P}{\sqrt{\left(\mu_P^{\rm{max}}\right)^2 + \left(I_P\alpha_P \right)^2}}, \,
\end{equation}
where $\mu_P^{\rm{max}}= 0.536~\rm{day^{-1}}$, and $I_P$ is the photosynthetically active radiation (PAR) which is calculated as
\begin{equation}\label{eq:PAR}
I_P=I_0\phi e^{-zK_w-\int_{0}^{z} K_{\rm{Chl}}~\rm{Chl}(\eta)~\rm{d}\eta}. \,
\end{equation}
Here, $\alpha_P =0.0538 ~\rm{day^{-1}~m^2~W^{-1}}$ is the initial slope of the photosynthesis-irradiance curve. $I_{0}$ is the total incoming solar radiation at the surface of water and $I_0 \phi$ is the photosynthetically available radiation with $\phi=0.43$. The chlorophyll (Chl) which is a surrogate
for phytoplankton biomass is estimated from the phytoplankton concentration.  $K_w = 0.059~\rm{m^{-1}}$ is the coefficient for the attenuation of light in water and $K_{\rm{Chl}} =
0.041~\rm{(mg)^{-1}}~\rm{m^2}$ is the coefficient for light attenuation due to  chlorophyll (\citet{BAGNIEWSKI[2011]}).

We have neglected  phytoplankton grazing by  zooplankton and nutrients  for the winter time simulation. A simulation that spans several seasons would require that equation  (\ref{eq:phy_eq}) be coupled to another set of equations governing nutrients and zooplankton.  
Since our simulations are limited to the winter when  nutrients are abundant and the grazing from zooplankton is considered insignificant, these equations are not incorporated (\citet{TAYLOR[2011a]}). 

\begin{figure}
	\centering	
	{\includegraphics[width=2.7in]{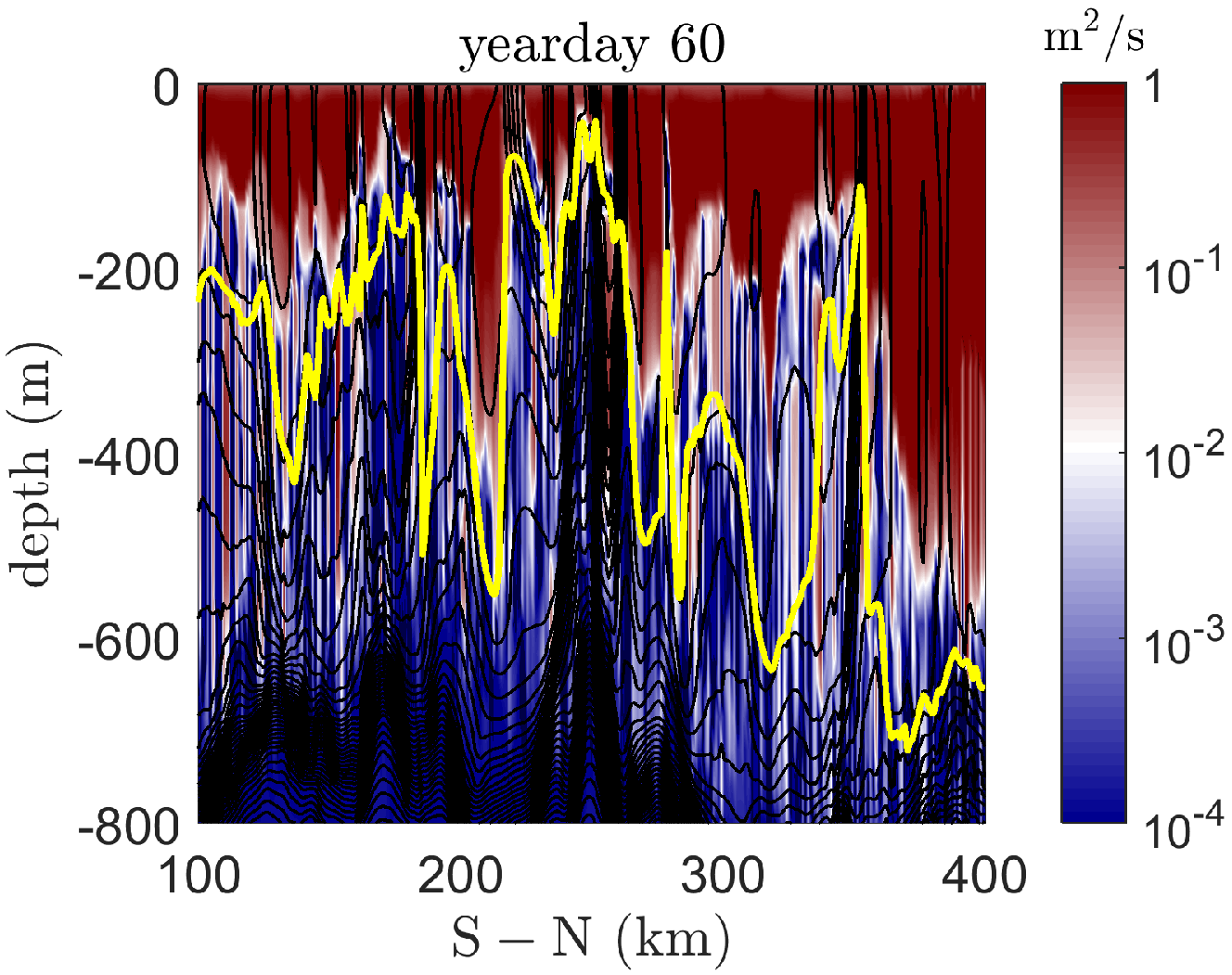}}
		{\includegraphics[width=2.7in]{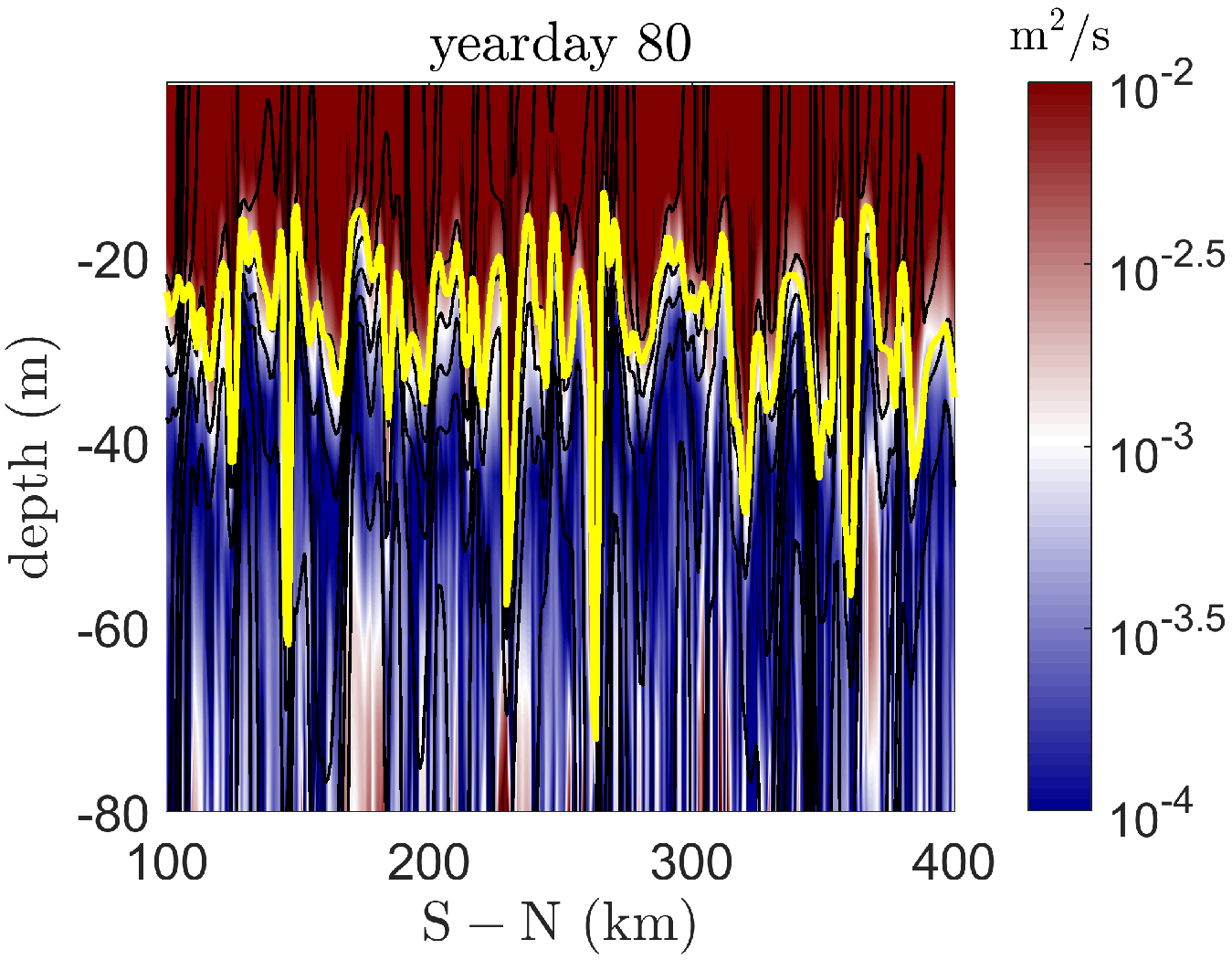}}
	\put(-370,150){(a)} \put(-175,150){(b)}\\ 
	\caption{The turbulent diffusivity ($K_z^d$) and the mixed layer in the meridional direction at (a) yearday 60; and  (b) yearday 80. The yellow line shows the MLD. Yearday 60 is a time of strong cooling with deep MLD and yearday 80 incurs a warming event with shallow MLD. In both cases, the region of active mixing agrees reasonably well with the MLD  although it is shallower  than the MLD  in some regions for strong cooling at yearday 60.}
	\label{fig:kz_mld}
\end{figure}

In our simulations, phytoplankton within the mixed layer are exposed to the light averaged over the MLD (i.e. $z \le$ MLD), which is determined as the depth where the density exceeds the density at the surface by 0.01 $\rm{kg/m^3}$. This assumption is justified by the fact that in the mixing layer, the turbulent mixing occurs on time scales much shorter than the required time scale for the growth of phytoplankton (\citet{MAHADEVAN[2012]}). The mixing layer is the region of the flow where the turbulent diffusivity ($K_z^d$) is high and hence turbulent mixing is vigorous. We test the suitability of this assumption by comparing the mixing layer depth with the MLD. 
The turbulent diffusivity ($K_z^d$) in the meridional direction in the middle of the zonal direction ($x=48$ km) is shown in Figure \ref{fig:kz_mld} for two different days (a) yearday 60 when there is a strong cooling event, strong mixing occurs  and the mixed layer is deep; and (b)  yearday 80 when the ocean surface is warming, the mixing is weak and hence the mixed layer is shallow. 
The MLD (yellow line) is plotted in both the figures.
It is clear that there is good agreement between the mixing layer with distinctively high turbulent diffusivity ($K_z^d$) and the mixed layer depth. This  confirms the suitability of our assumption for mixing the light and averaging it in the mixed layer. 

\section{Results and discussion}\label{sec:results}
In this section, we address the three main questions raised in section \ref{sec:intro}.

\subsection{Hourly vs. weekly-averaged forcing}\label{sec:flux}
\begin{figure}
	\centering
	
	{\includegraphics[width=2.5in]{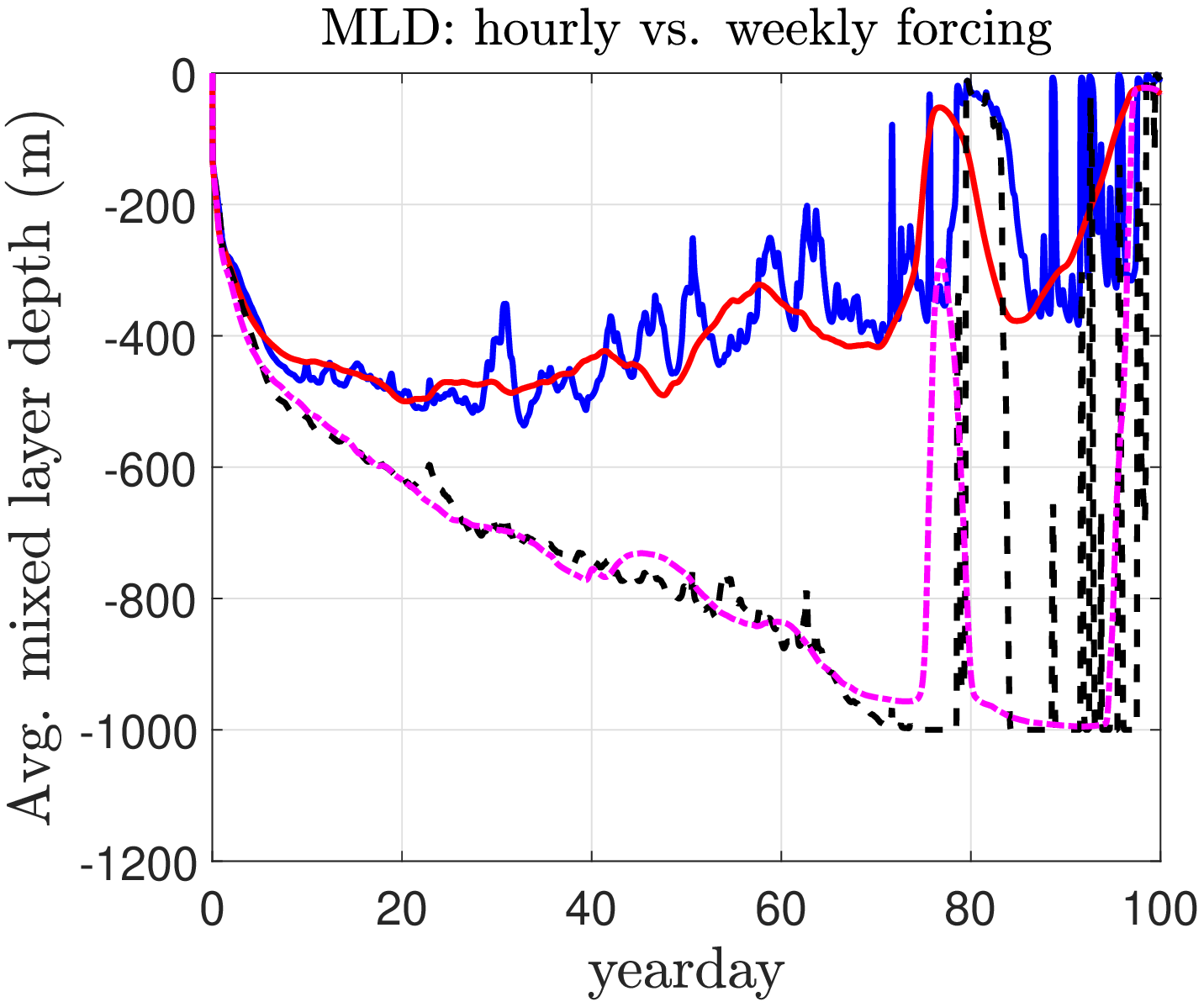}}
	\put(-160,145){(a)}
	{\includegraphics[width=2.5in]{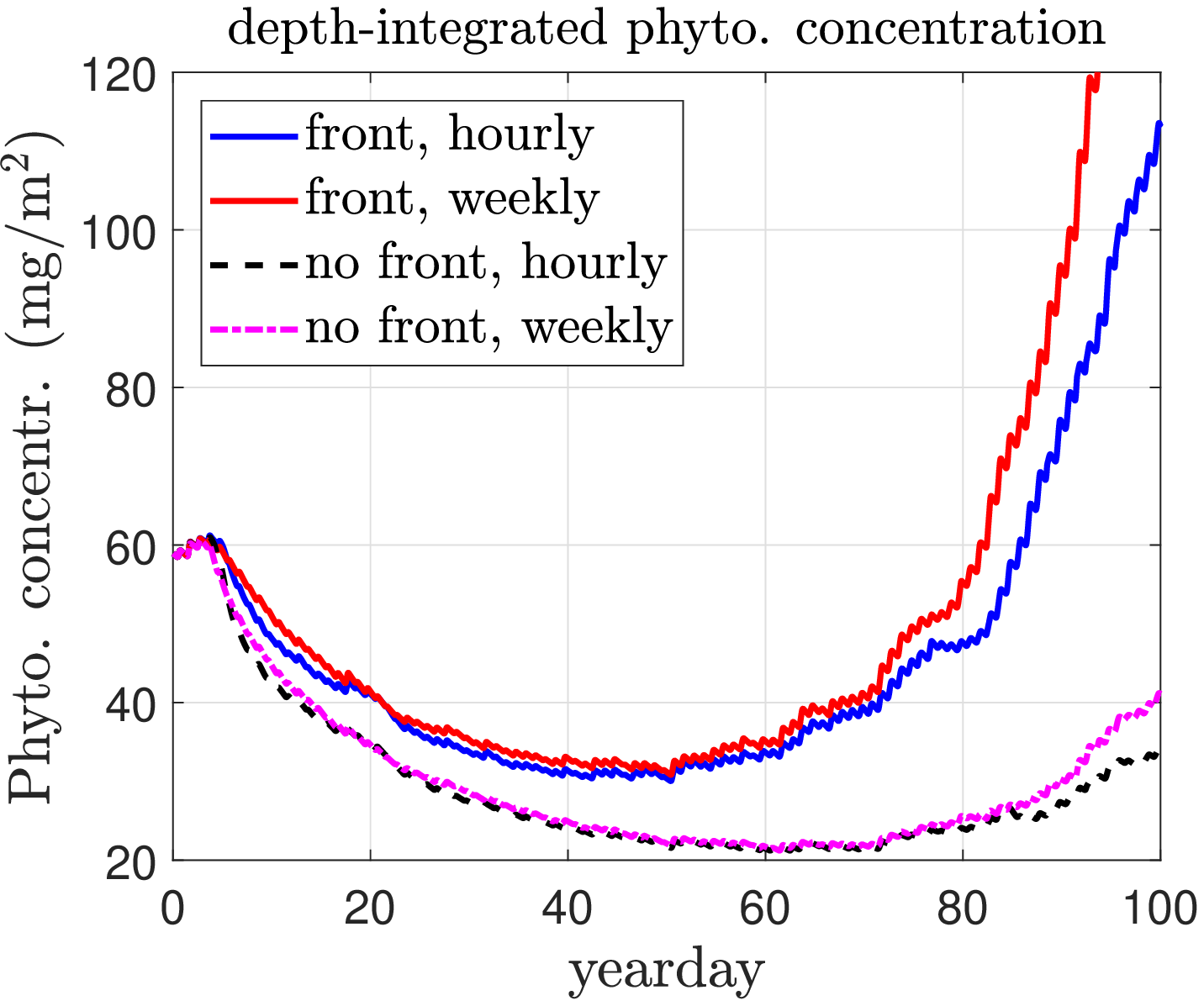}}
	\put(-160,145){(b)} 
	\caption{The comparison of hourly forcing and weekly-averaged forcing effect on (a) the average mixed layer depth; and (b) the depth-integrated phytoplankton concentration. The cases without fronts show substantially deeper mixed layer depths and lower phytoplankton concentration compared to the cases with fronts.}
	\label{fig:mld_phy_avg}
\end{figure}

The change in average MLD with time and the depth-integrated phytoplankton down to $z \approx -400$ m are shown in Figure \ref{fig:mld_phy_avg} for hourly and weekly-averaged forcing. These quantities are averaged zonally as well as from $y=100-400~\rm{km}$. 
In the  rest of the paper these quantities are calculated as described here. 
Also, to better illustrate the effect of forcing on the MLD evolution and its influence on  phytoplankton growth in the winter, two additional numerical simulations are implemented without fronts and with the same hourly and averaged fluxes. 

Due to intermittency of the hourly $Q_{\rm{net}}$ and wind stress, the mixed layer stratification changes within days (Figure \ref{fig:mld_phy_avg}a), particularly when fronts exist.
However, comparison of the average phytoplankton concentration from hourly and weekly-averaged fluxes in Figure \ref{fig:mld_phy_avg}b does not show a significant difference. This surprising result  is in contrast to the expectation that when MLD shoals in the winter, the phytoplankton production should increase accordingly. 

\begin{figure}
	\centering 	
	{\includegraphics[width=3.in]{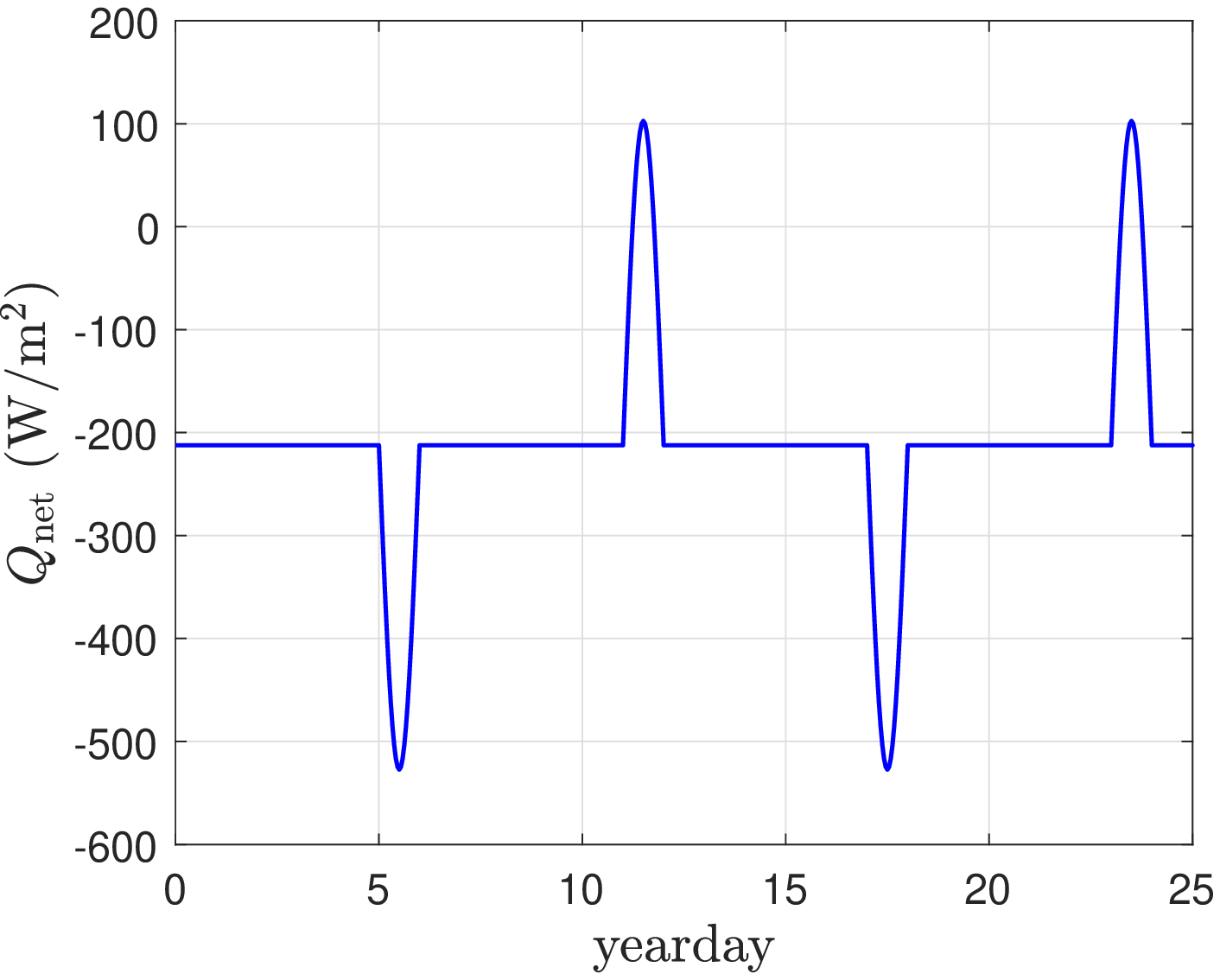}}
	\put(-190,165){(a)}\\
	{\includegraphics[width=2.5in]{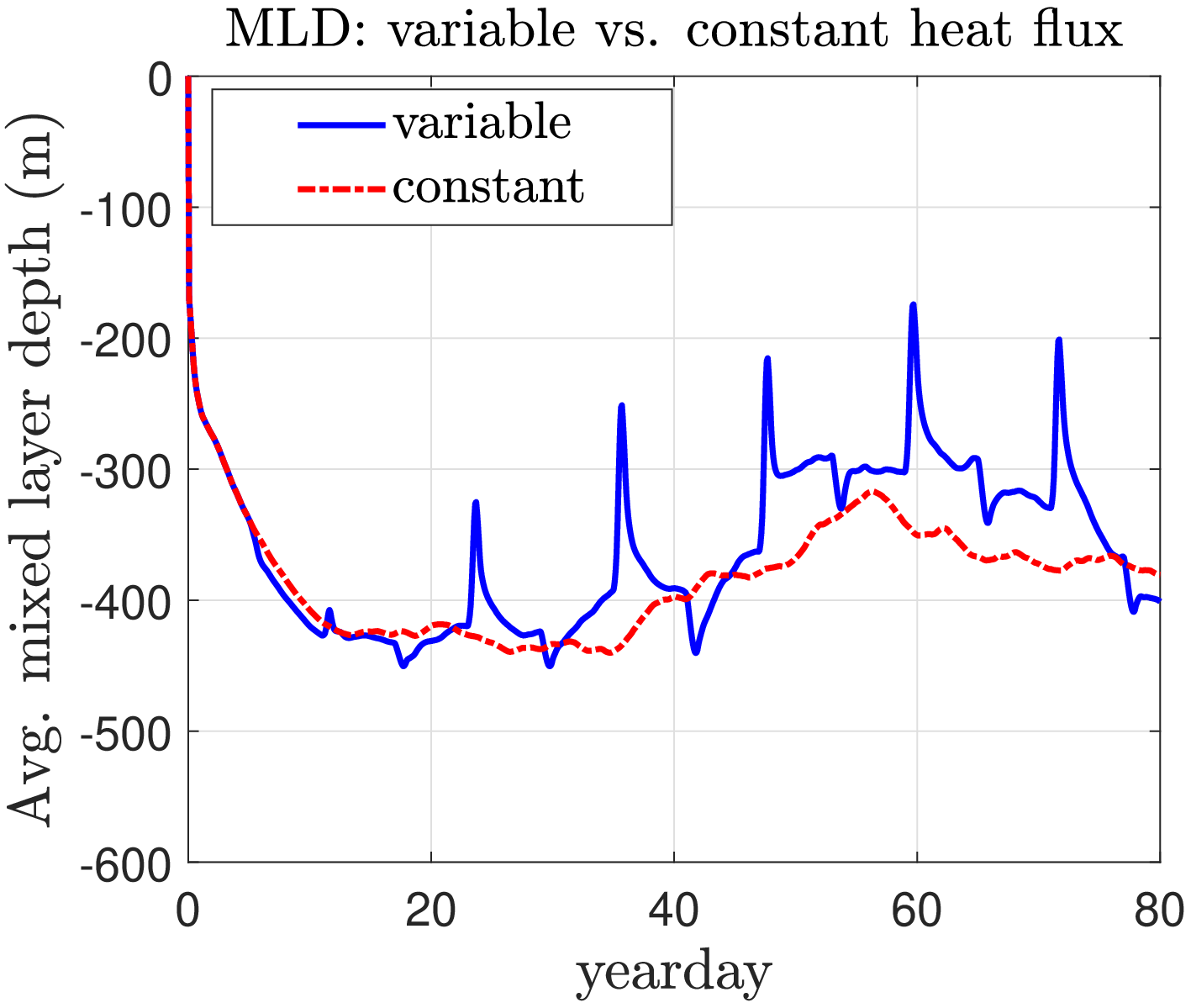}}
	\put(-160,145){(b)}
	{\includegraphics[width=2.5in]{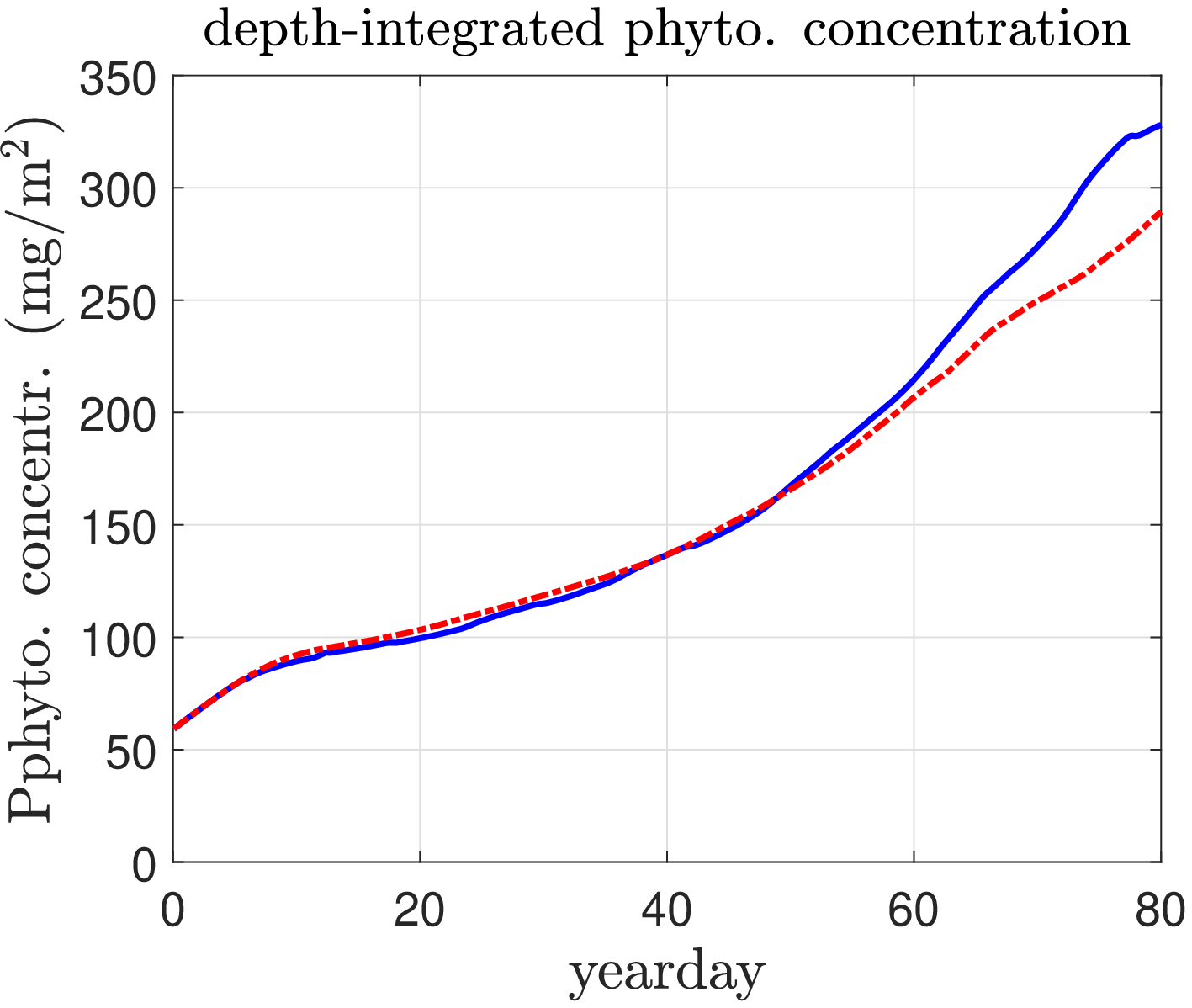}}
	\put(-160,145){(c)} 
	\caption{Response of the mixed layer and depth average phytoplankton concentration to synthetically varying net heat flux. (a) Synthetically varying net heat flux; (b) mixed layer response; and (c) the average phytoplankton concentration (c). Each cooling or warming event has a one-day duration and an event occurs every 5 day.} 
	\label{fig:mld_phy_var_const}
\end{figure}

For further assessment of effects of the forcing on phytoplankton production, we carried out a simulation  
with constant forcing $Q_s=47.97~\rm{W/m^2}$, $Q_{\rm{net}}=212.31~\rm{W/m^2}$, $\tau_x=0.158~\rm{N/m^2}$ and $\tau_y= 0.025~\rm{N/m^2}$, which are the averaged fluxes for the first 100 days of 2008 as also shown in Tables \ref{tab:forcing_cool} and \ref{tab:forcing_wind}. A second case was simulated similar to the first scenario with constant wind except that there was a warming or cooling event every 5 days as shown in Figure \ref{fig:mld_phy_var_const}a, with the same average $Q_{\rm{net}}$ as in the first scenario. Each warming or cooling event lasted for one day, varying sinusoidally during the time of occurrence with an amplitude of 315 $\rm{W/m^2}$, yielding a standard deviation of ${\sim 65}~\rm{W/m^2}$ for $Q_{\rm{net}}$ that is comparable to the standard deviation of the actual $Q_{\rm{net}}$ presented in Table \ref{tab:forcing_cool}. The time-varying $Q_{\rm{net}}$ in Figure \ref{fig:mld_phy_var_const}a only shows the first 25 days for better illustration of the net heat flux ( $Q_{\rm{net}}$). 

 %

The results  show that the mixed layer changes much more due to warming than cooling (Figure \ref{fig:mld_phy_var_const}b). On the other hand, the results of Figure \ref{fig:mld_phy_var_const}c prove that the phytoplankton concentration for both cases are very similar for the first 70 days. After yearday 70, phytoplankton concentration from the variable net heat flux becomes more than the constant forcing. A comparison of the minimum and maximum phytoplankton concentration (not shown here) showed similar behaviour. 
The results highlight the fact that episodic heat fluxes during the winter has relatively little effect on the average phytoplankton concentration in the North Atlantic.

 In a related work, \citet{WHITT[2017]} studied the effect of intermittent winds in nutrient-limited conditions. They observed an increase of nutrients and hence more phytoplankton production near fronts due to oscillations in wind. In contrast, we assume nutrients are abundant in the winter implying that the only parameter controlling the growth of  phytoplankton is the sunlight. The current study shows that intermittency in MLD due to fluxes has little net effect on average phytoplankton growth in the winter. 

It is also worth noting that phytoplankton mostly grow in these synthetic simulations unlike previously presented results in Figure \ref{fig:mld_phy_avg}b. The persistent increase in phytoplankton concentration is due to the constant shortwave during the whole simulation which consequently implies that PAR is constantly available. In the real ocean, the low-level shortwave is available only for a fraction of the day in the winter. Therefore, although stratification can improve conditions for the growth of phytoplankton, the limitation in light availability during the winter prevents phytoplankton from leveraging the intermittent stratification.  


\begin{figure}[ht!]
	\centering 	
	{\includegraphics[width=3.in]{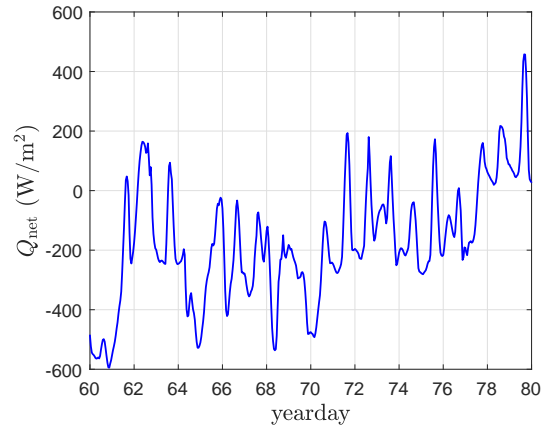}}
	\caption{The neat heat flux $(Q_{\rm{net}})$ for yeardays $60 - 80$. Most warming events last less than a day which are much shorter than the growth time scale of phytoplankton.}
	\label{fig:q_net_short}
\end{figure}
  
\begin{figure}[ht!]
	\centering 	
	{\includegraphics[width=3.in]{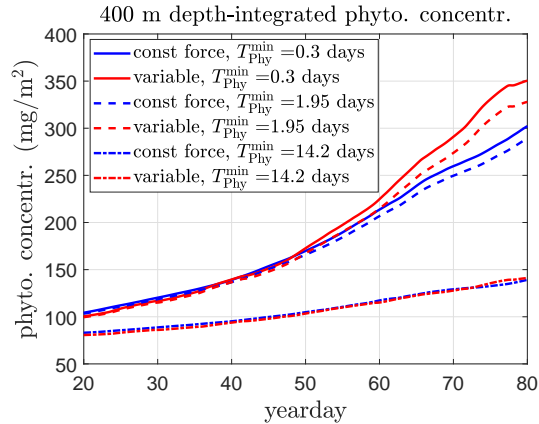}}
	\caption{The average phytoplankton concentration for different growth time scales and forcing.}
	\label{fig:var_mu}
\end{figure}

The relative insensitivity of phytoplankton to the high episodic fluxes warrants an examination of the time scales pertinent to phytoplankton growth. We can consider a growth time scale for phytoplankton  $T_{\rm{Phy}}=1/ (\mu_P-m_P)$. The smallest corresponding time scale of the model we employed is $T^{\rm{min}}_{\rm{Phy}}=1/(\mu_P^{\rm{max}}-m_p)\approx 1.95$ days. Moreover, examining the net heat flux ($Q_{\rm{net}}$) for yeardays $60 - 80$ presented in Figure \ref{fig:q_net_short}, except after yearday $\sim$77, most  positive net heat fluxes ($Q_{\rm{net}}$) last a fraction of a day. Consequently, the duration of ML shoaling is also a fraction of a day. Therefore, before phytoplankton have sufficient time to grow, cooling occurs. To test the effect of phytoplankton growth time scale, we performed simulations with similar conditions used for simulations in Figure \ref{fig:mld_phy_var_const}, but for several phytoplankton growth time scales ranging from $0.3-14.2$ days. Results  for three time scales $T^{\rm{min}}_{\rm{Phy}}=0.3, 1.95$ and 14.2 (Figure \ref{fig:var_mu}) show that   as we expected, for the long phytoplankton growth time scale of $T^{\rm{min}}_{\rm{Phy}}=14.2$ days the average phytoplankton concentration for constant forcing and variable forcing are very similar. With decrease of $T_{\rm{Phy}}$, phytoplankton concentration increases for both forcing conditions and the difference in phytoplankton concentration between variable forcing and constant forcing grows. However, the difference between the phytoplankton concentration for the constant and variable forcing for each of these time scales is not significant.

To better understand the reason for insensitivity of phytoplankton growth to intermittent air-sea fluxes, we need to take into account the role of MLD. In Figure \ref{fig:mld_phy_var_const}b, the average MLD is about $300-400$ m. In spite of the intermittent shoaling due to variable air-sea fluxes, the average MLD transiently shoals to ~200 m. Considering that ML is very deep in winter, phytoplankton hardly get transported to the euphotic layer for sufficient amount of time to grow. Hence, when forcing is intermittent, phytoplankton do not have considerably better conditions for growth than when the forcing is not intermittent.

   \begin{figure}[]
	\centering
	{\includegraphics[width=4in]{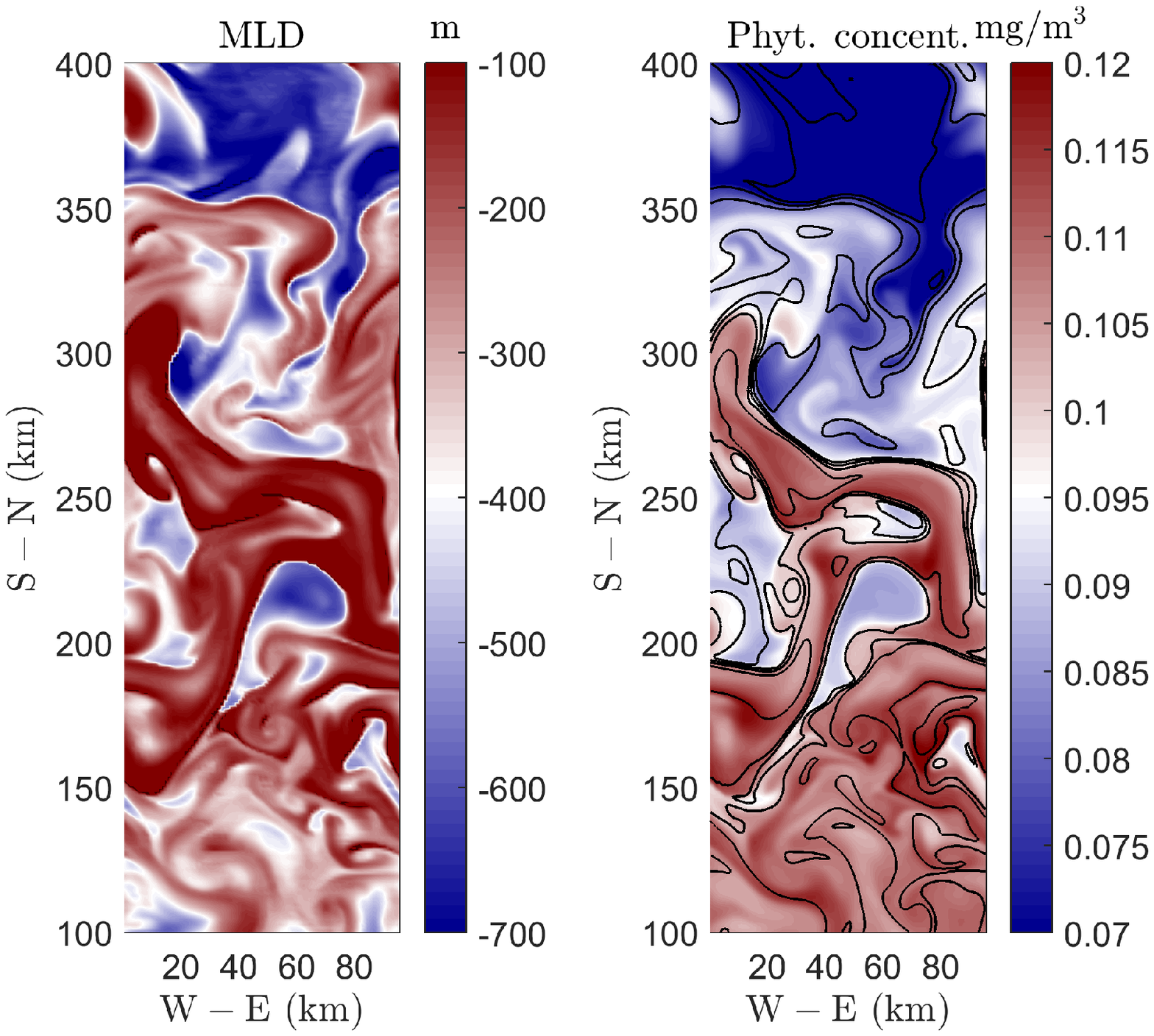}}
	\put(-248,230){(a)}    	\put(-120,230){(b)}\\
	{\includegraphics[width=3in]{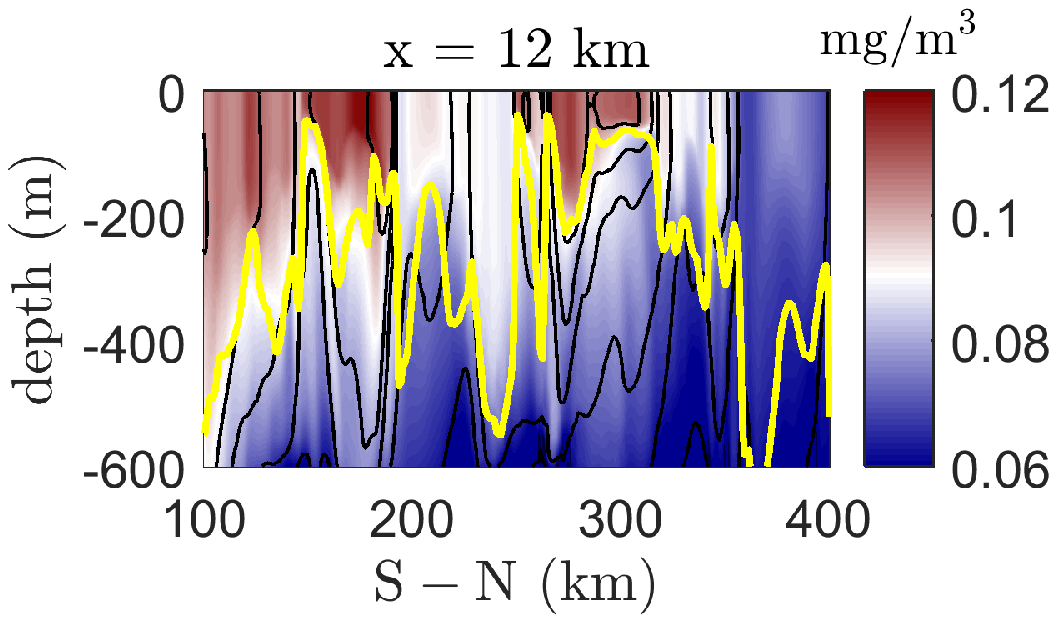}}
	\put(-190,130){(c)}    	\\
	{\includegraphics[width=5.5in]{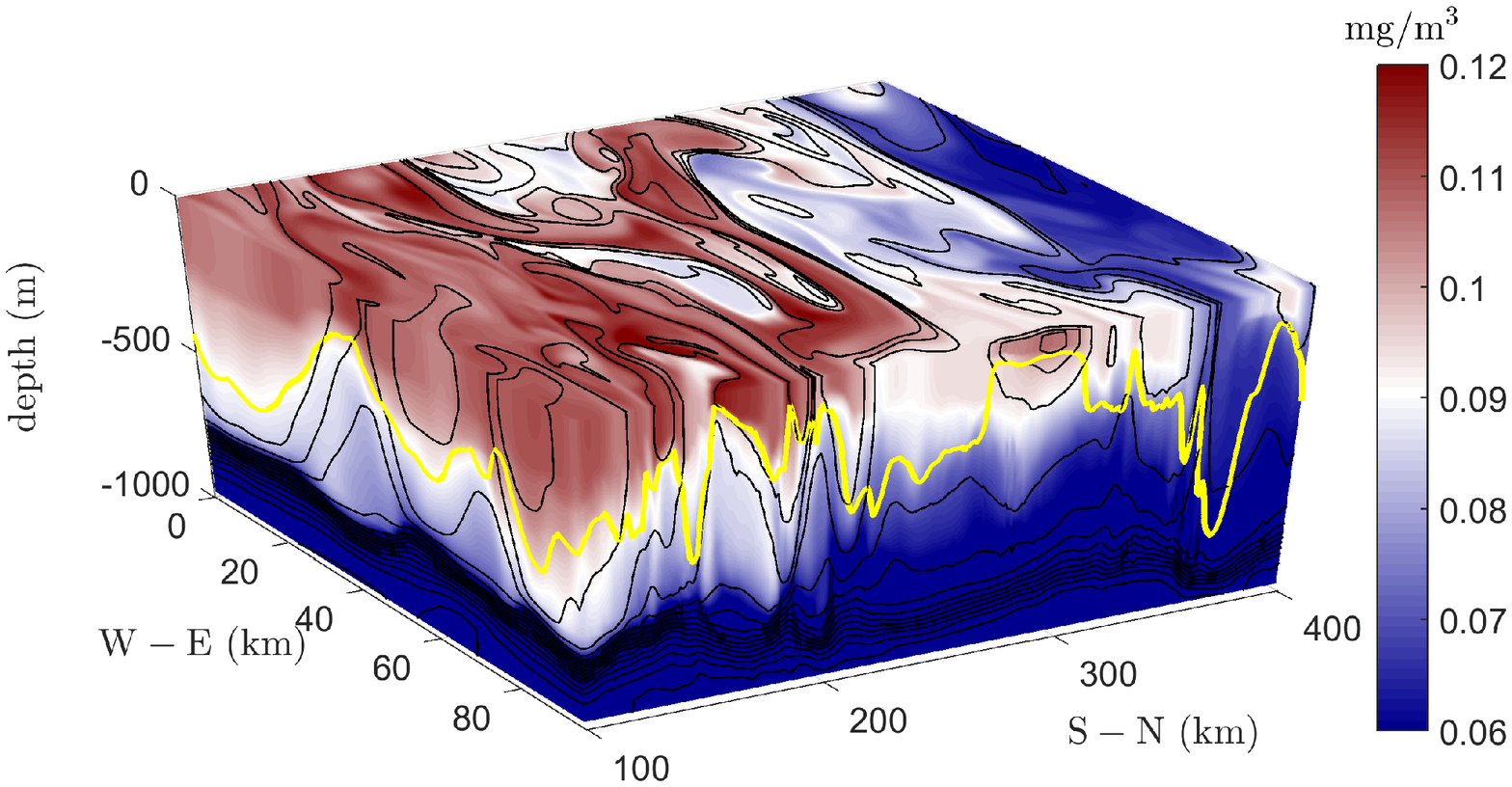}}
	\put(-380,170){(d)}   
	\caption{The simulation results at yearday 60 at midnight  showing (a) the mixed layer depth; (b) the concentration of the phytoplankton on the horizontal surface at $z\approx-5.85~\rm{m}$; (c) the phytoplankton concentration in the meridional direction at $x=12$ km; (d) the phytoplankton concentration at $x=$ 96 km, $y=$ 100 km and the topmost boundary of the domain. The yellow solid line in Figures c and d shows the MLD.}
	\label{fig:phy_mld_xy_hour}
\end{figure}

\subsection{Effect of fronts}\label{sec:front_eff}
In the previous section we showed that in the short daylight hours of the winter, sudden changes in air-sea fluxes do not considerably contribute to the growth of phytoplankton. However, Figure \ref{fig:mld_phy_avg}a shows that due to the slumping of  fronts and the consequent restratification, the average mixed layer depth in cases with fronts  is substantially shallower than the cases with no horizontal gradient of the density for time scales much longer than $T_{\rm{Phy}}$. Shallower mixed layers resulting from frontal instabilities increase phytoplankton residence time in the euphotic layer, resulting in their growth (\cite{MAHADEVAN[2012]}).  
Comparing Figures \ref{fig:mld_phy_avg}a and \ref{fig:mld_phy_avg}b, we observe  fronts result in shallow mixed layers for both hourly and averaged fluxes and therefore enhanced production compared to cases with no front during the winter. Hence, fronts enormously contribute to the production and sustenance of  phytoplankton in winter months.
The role of fronts has been previously appreciated in studies on the spring bloom showing that the restratification due to fronts is a key parameter for the early initiation of the spring bloom of phytoplankton (\citet{MAHADEVAN[2012]}).

So far we have focused on the variation of the average MLD and phytoplankton. Now, we briefly evaluate how  phytoplankton and the MLD  change spatially when fronts are present. To that end, we show the MLD as well as the phytoplankton concentration at $z\approx -5.85~\rm{m}$ in Figures \ref{fig:phy_mld_xy_hour}a and b. The results we show here are for yearday 60 at midnight and do not change significantly for other times of the day.
We can see that there is high spatial variability in MLD and phytoplankton. Also, the distribution of the MLD agrees fairly well with the distribution pattern of  phytoplankton  in the domain such that the phytoplankton concentration is normally more where the MLD is shallow and  low  where the MLD is  deep, leading  to patches of phytoplankton in the domain. The same agreement between MLD and phytoplankton concentration can also be seen in Figures \ref{fig:phy_mld_xy_hour}c and d. Figure \ref{fig:phy_mld_xy_hour}c shows the phytoplankton concentration and MLD in the meridional direction at $x=12$ km, where the yellow line is the mixed layer depth and Figure \ref{fig:phy_mld_xy_hour}d  is a three-dimensional representation of the phytoplankton concentration as well as the MLD (yellow line).  

In spite of relatively good agreement between MLD and phytoplankton concentration, it is clear that especially near the south, the mixed layer is deep while the phytoplankton concentration is high. In our three dimensional simulations, phytoplankton get advected between different regions of the flow. A mechanism that contributes to advection is the Ekman transport that due to the dominant down-front, eastward wind in our simulation, transports phytoplankton towards the south. Also, as the phytoplankton mortality  time scale is large, advected phytoplankton do not decrease quickly in the south.
A Lagrangian study  could  reveal better how phytoplankton evolve in different phases of the day and in different places in the domain, which is beyond the scope of the current study.  
In the next section, we will discuss the effect of numerical resolution on the variability of the MLD and phytoplankton and highlight the importance of resolving submesoscale fronts.

\subsection{Effect of frontal strength and spatial resolution}\label{sec:reso}
The results discussed so far highlight the undeniable effect of fronts on the production and sustenance of the phytoplankton population in the winter. This means that  modeling phytoplankton in the winter requires resolving the fronts and their related  instabilities that lead to stratification. 
The criterion for determining the grid resolution is the Rossby radius of deformation ($L_R$), which is defined as 
\begin{equation}\label{eq:LR}
L_R = \frac{N H}{f}, \,
\end{equation}
where $f$ is the Coriolis frequency and $H$ is the relevant water depth (i.e. the mixed layer depth).
The Rossby radius of deformation, which essentially shows the competition of buoyancy and rotational forces (i.e. larger Rossby radius means that buoyancy forces are more dominant) is the threshold above which the flow starts to become geostrophic. In order to resolve submesoscale eddies, which occur at scales where effects of geostrophic balance diminish, the spatial resolution has to be sufficiently smaller than $L_R$. Considering initial conditions shown in Figure \ref{fig:init_den}b, we can estimate the $L_R$ for our studied case as
\begin{equation}\label{eq:LR_cal}
L_R =\frac{ N H}{f} \sim \frac{\left(\sim 1\times 10^{-3}~(\rm{s^{-1}}) \right) \times \sim 300~ \rm{(m)}}{1.114 \times10^{-4}~ \rm{(s^{-1})}} \sim 2700~ \rm{m}, \,
\end{equation}
where $f = 1.114 \times10^{-4}~\rm{s^{-1}}$ is the Coriolis (inertial) frequency at $50^{\circ}\rm{N}$. Therefore, the spatial resolution for capturing submesoscale processes needs to be at least $\Delta x, \Delta y \sim 1/4 \times 2700 ~\rm{m}$ to resolve an instability wave length. The simulations discussed thus far,  for which the spatial resolution was 1 km, can marginally resolve submesoscale processes. We next address the impact of spatial resolution.

\begin{figure}[ht!]	
	\centering
	{\includegraphics[width=2.7in]{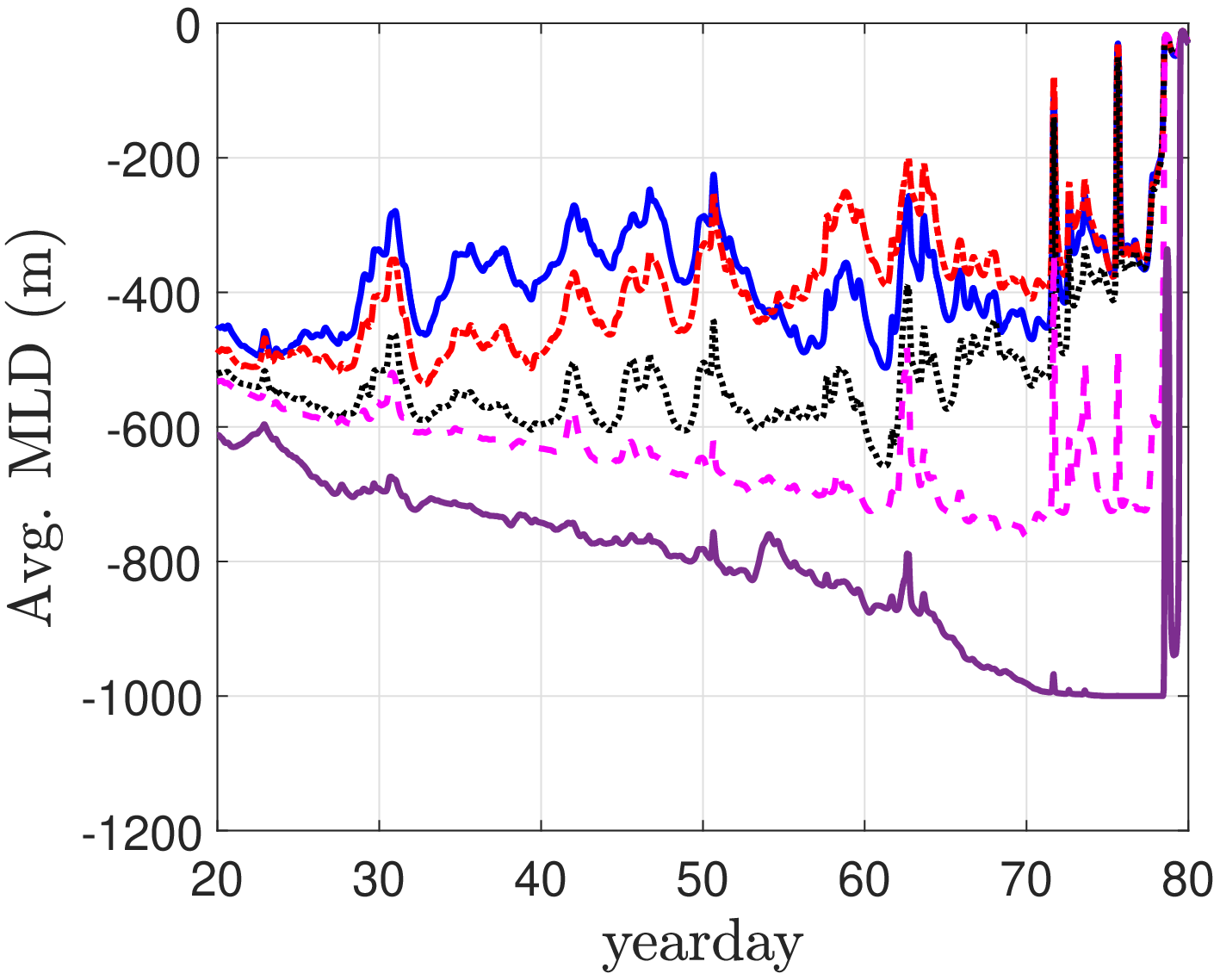}}
	{\includegraphics[width=2.7in]{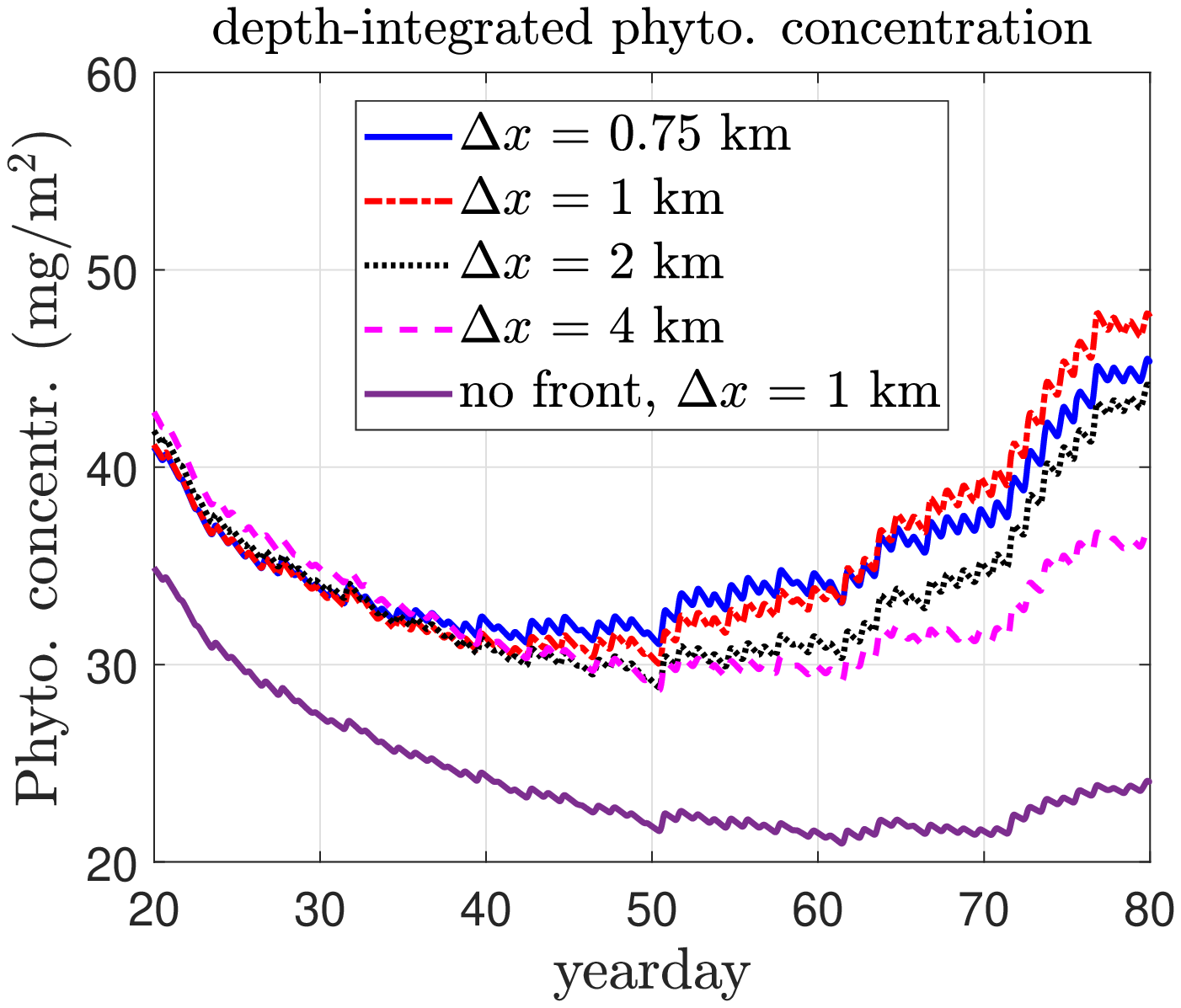}}
	\put(-400,160){(a)}  	 	\put(-200,160){(b)}\\
	{\includegraphics[width=2.7in]{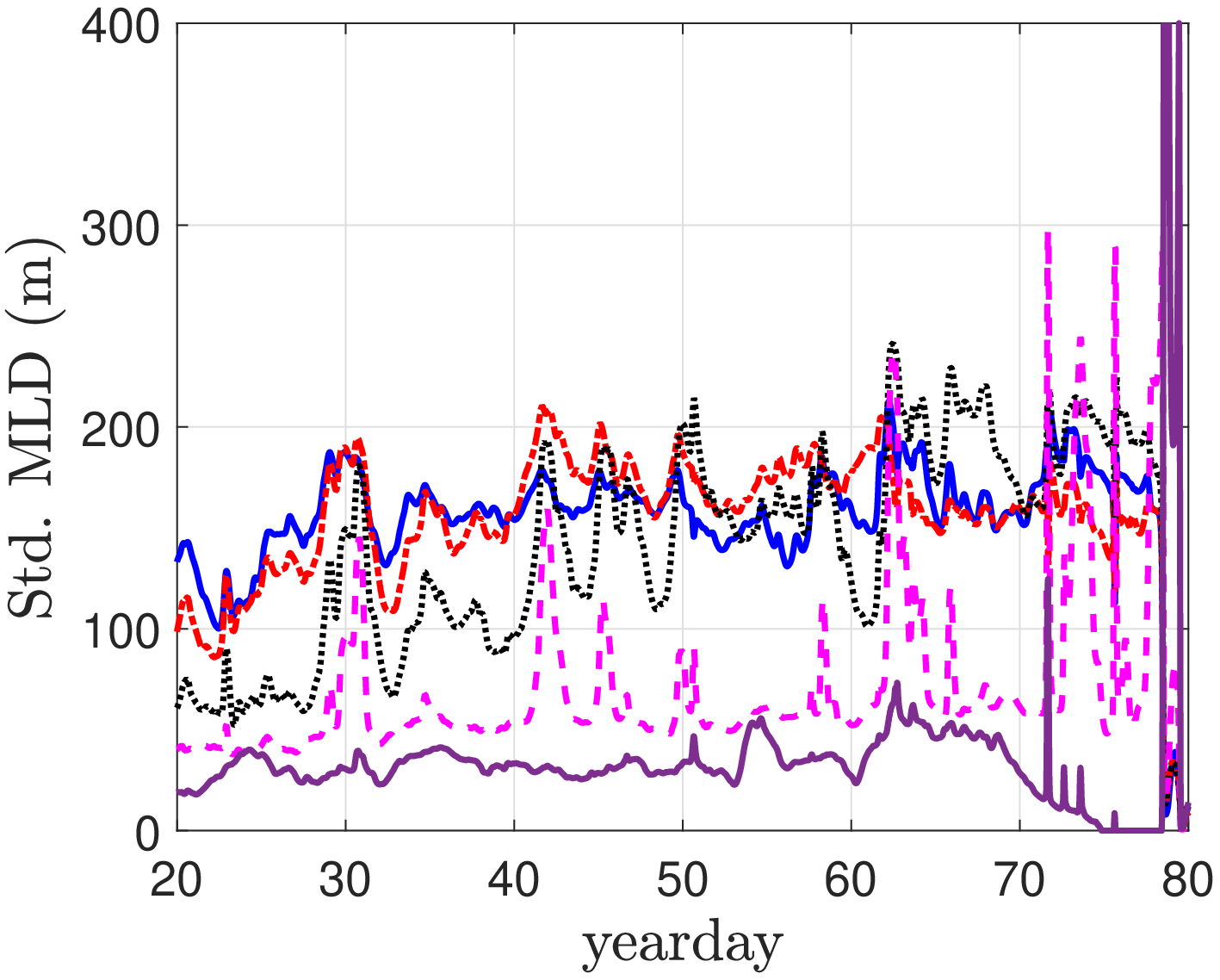}}
	{\includegraphics[width=2.7in]{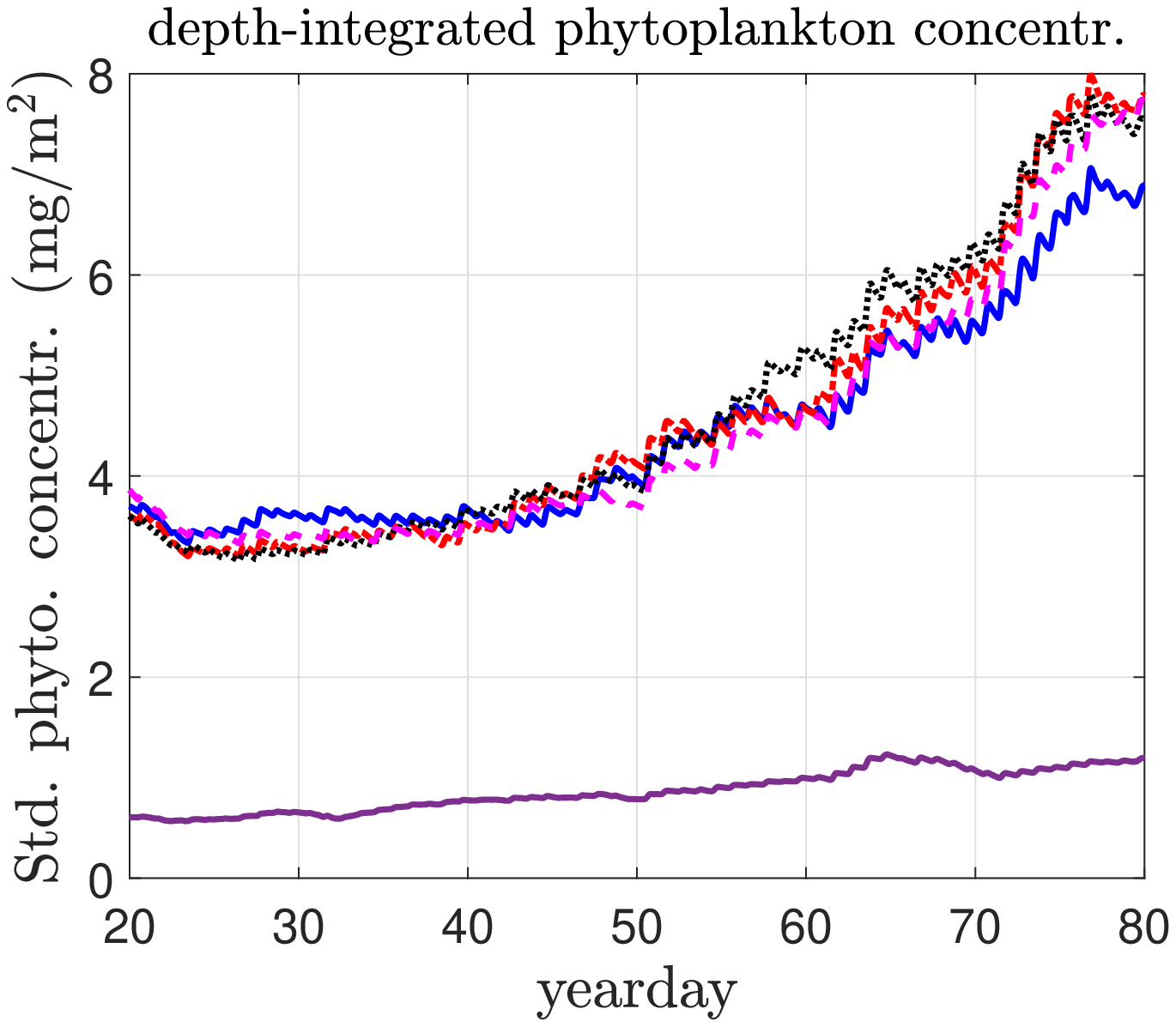}}
	\put(-400,160){(c)}  	 	\put(-200,160){(d)}
	\caption{ (a) Comparison of the average MLD; (b) the depth-integrated phytoplankton concentration; (c) the standard deviation of the MLD; (d) the standard deviation of the depth-integrated phytoplankton concentration; for the grid resolution of $0.75,~1,~2$ and $4$ km and no front case with 1 km resolution.}
	\label{fig:MLD_phy_num_res}
\end{figure}

\begin{figure}[]	
	\centering	
	{\includegraphics[width=2.70in]{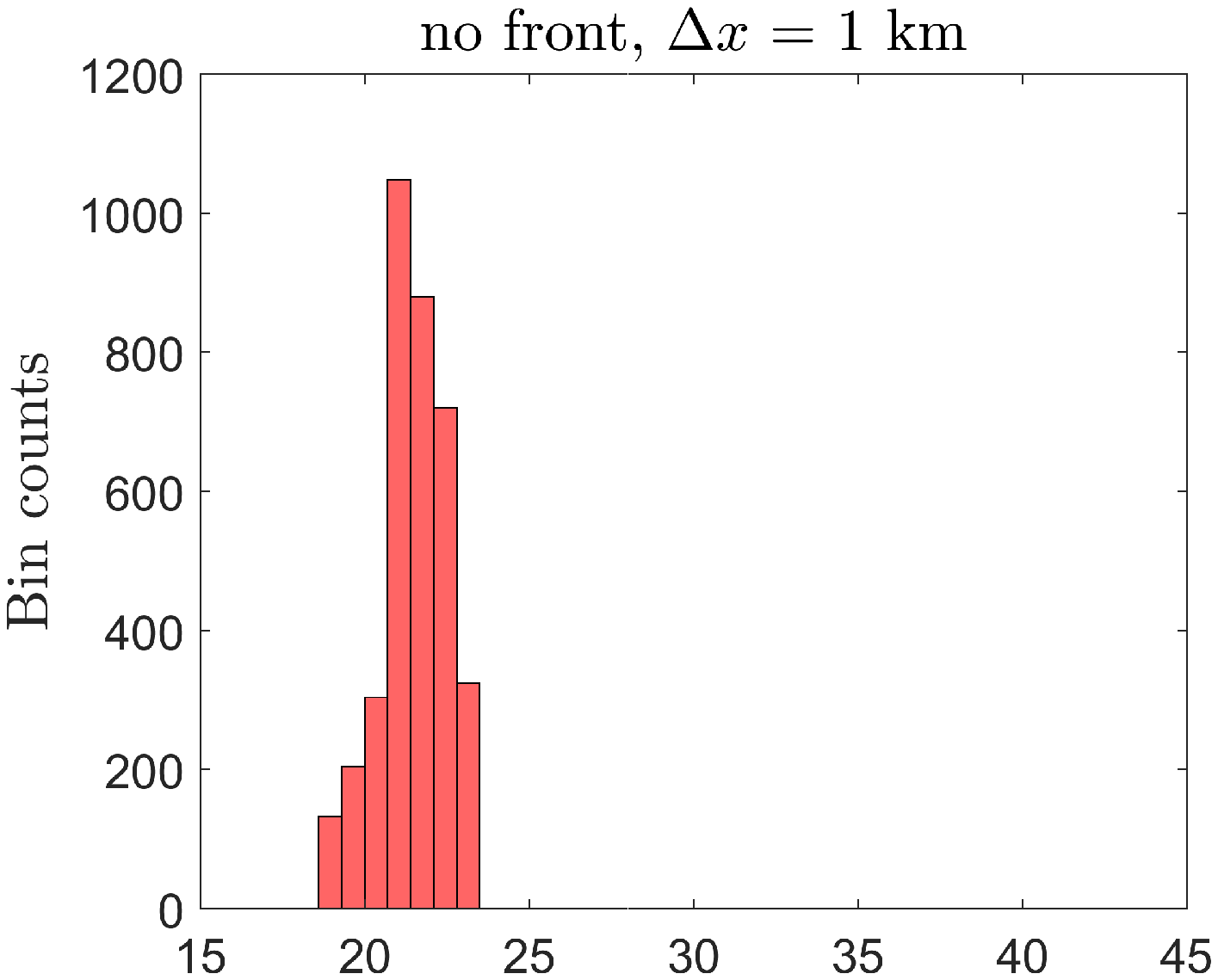}}
		{\includegraphics[width=2.70in]{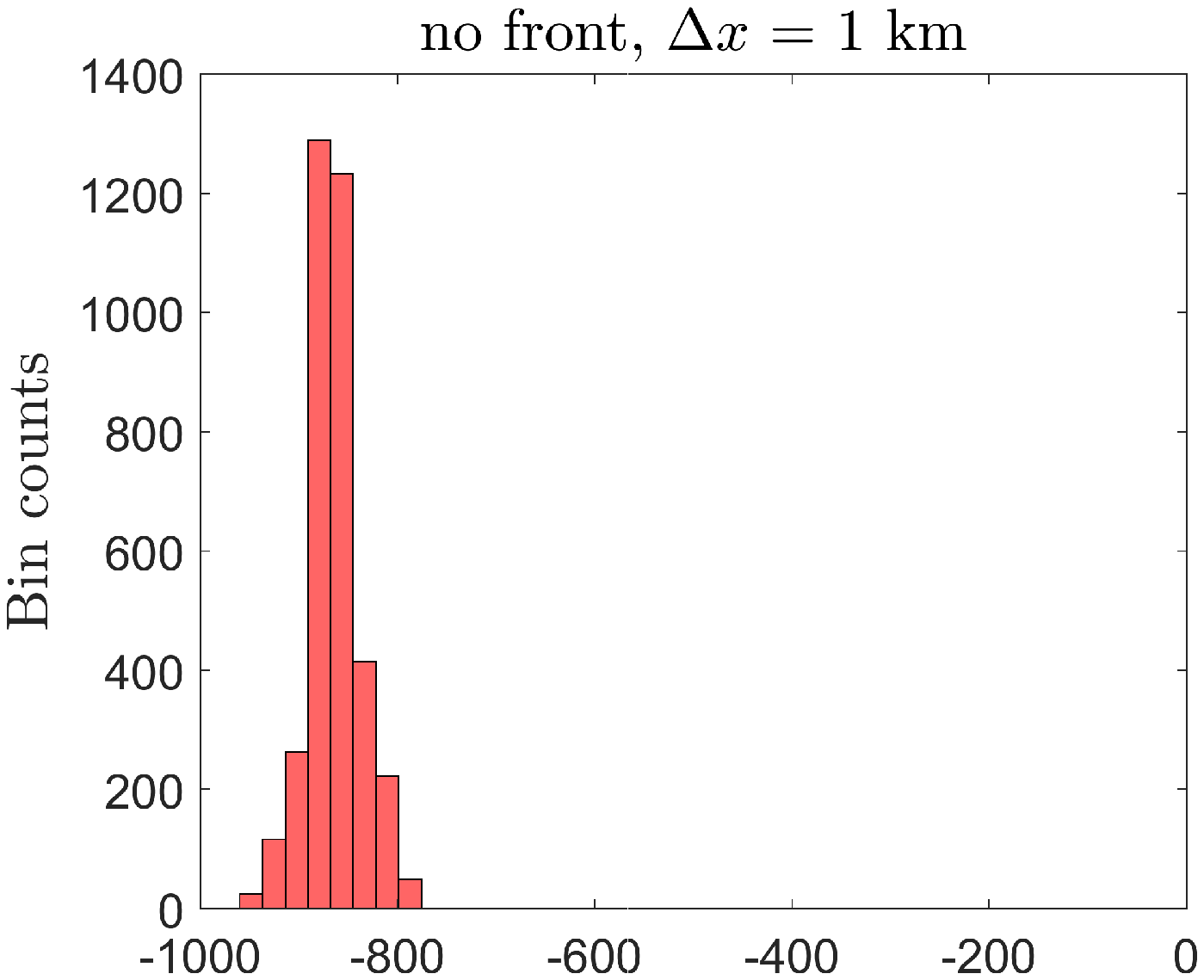}}
	\put(-400,145){(a)} \put(-200,145){(b)} \\
		{\includegraphics[width=2.70in]{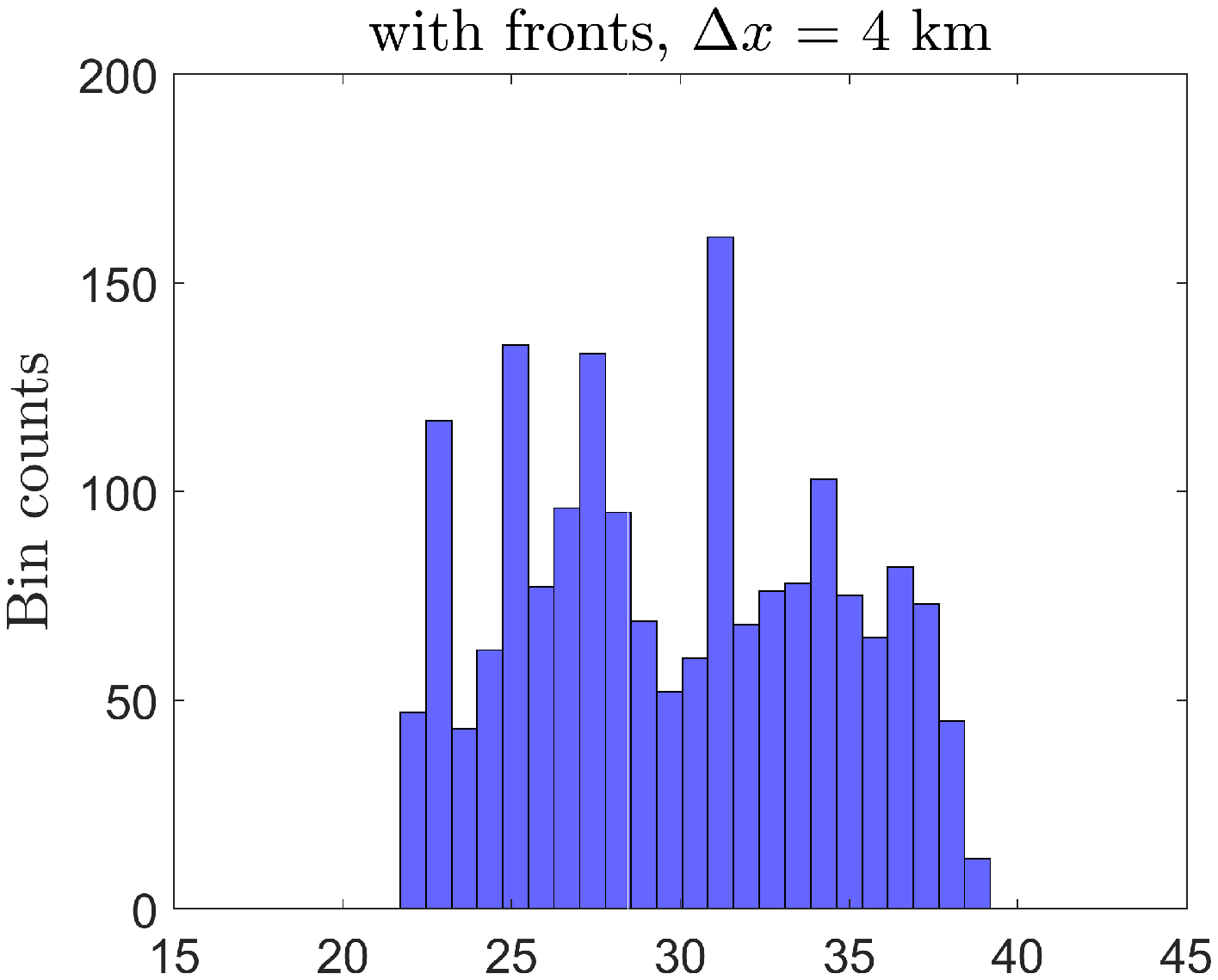}}
		{\includegraphics[width=2.70in]{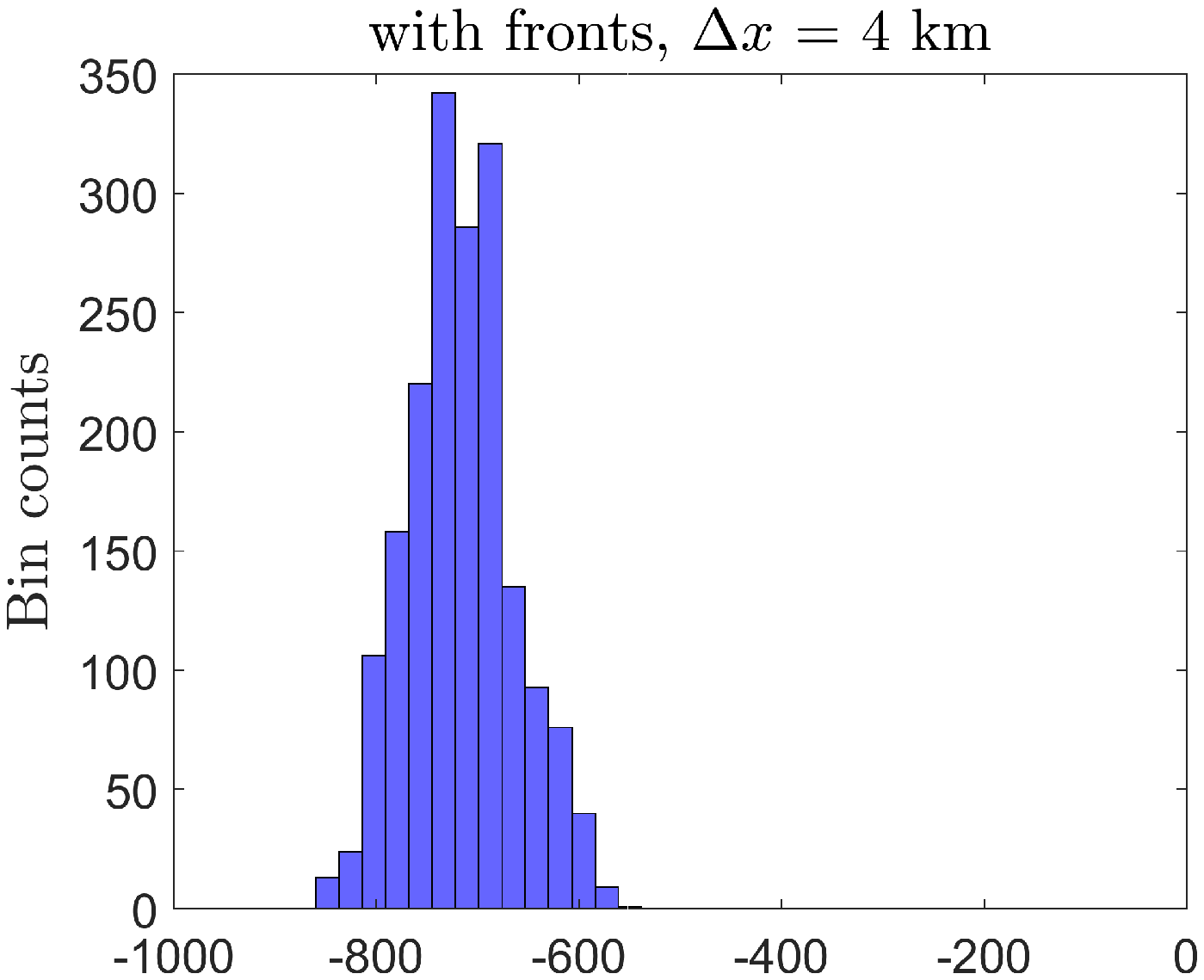}}
			\put(-400,145){(c)} \put(-200,145){(d)} \\		
	{\includegraphics[width=2.7in]{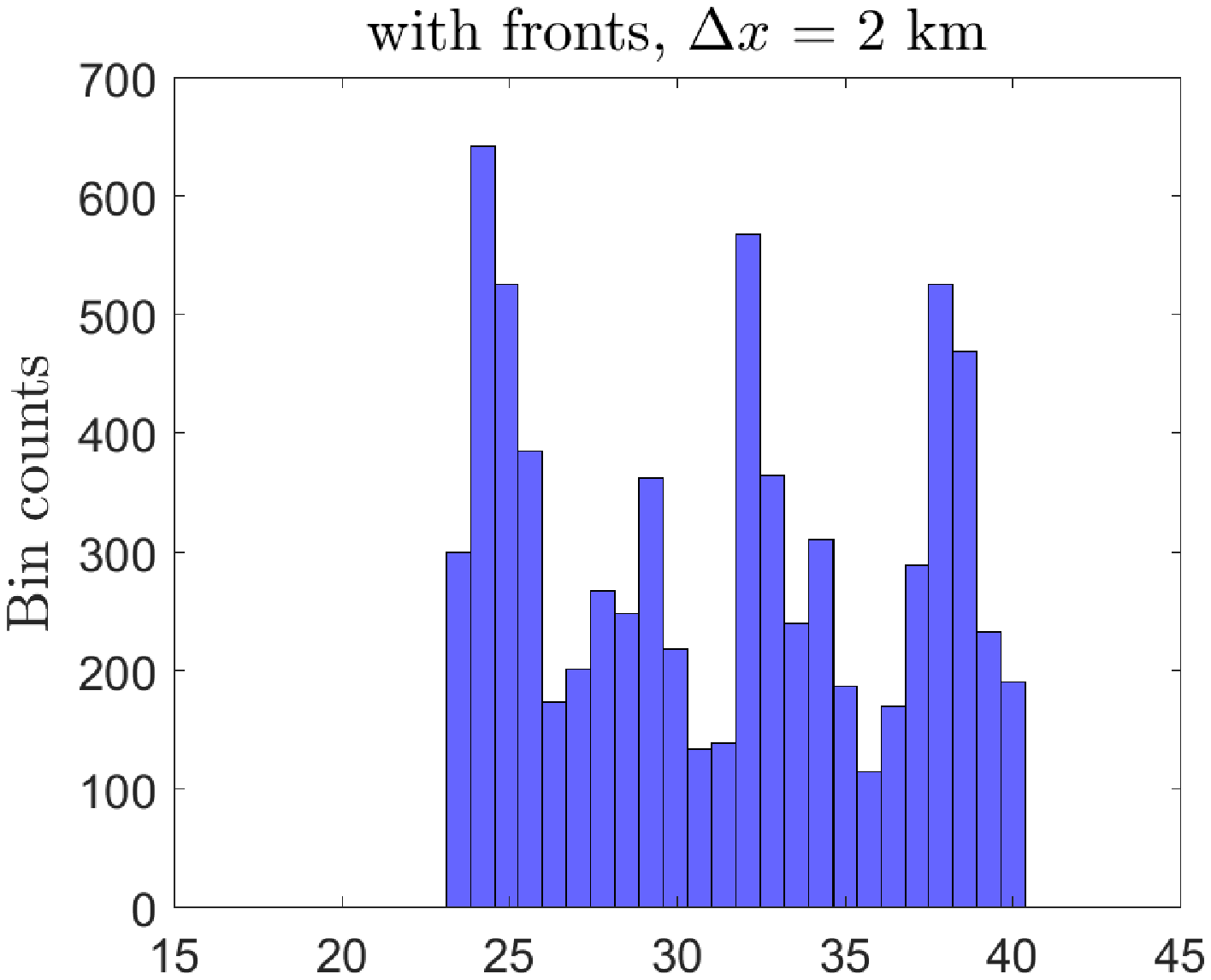}}
	{\includegraphics[width=2.7in]{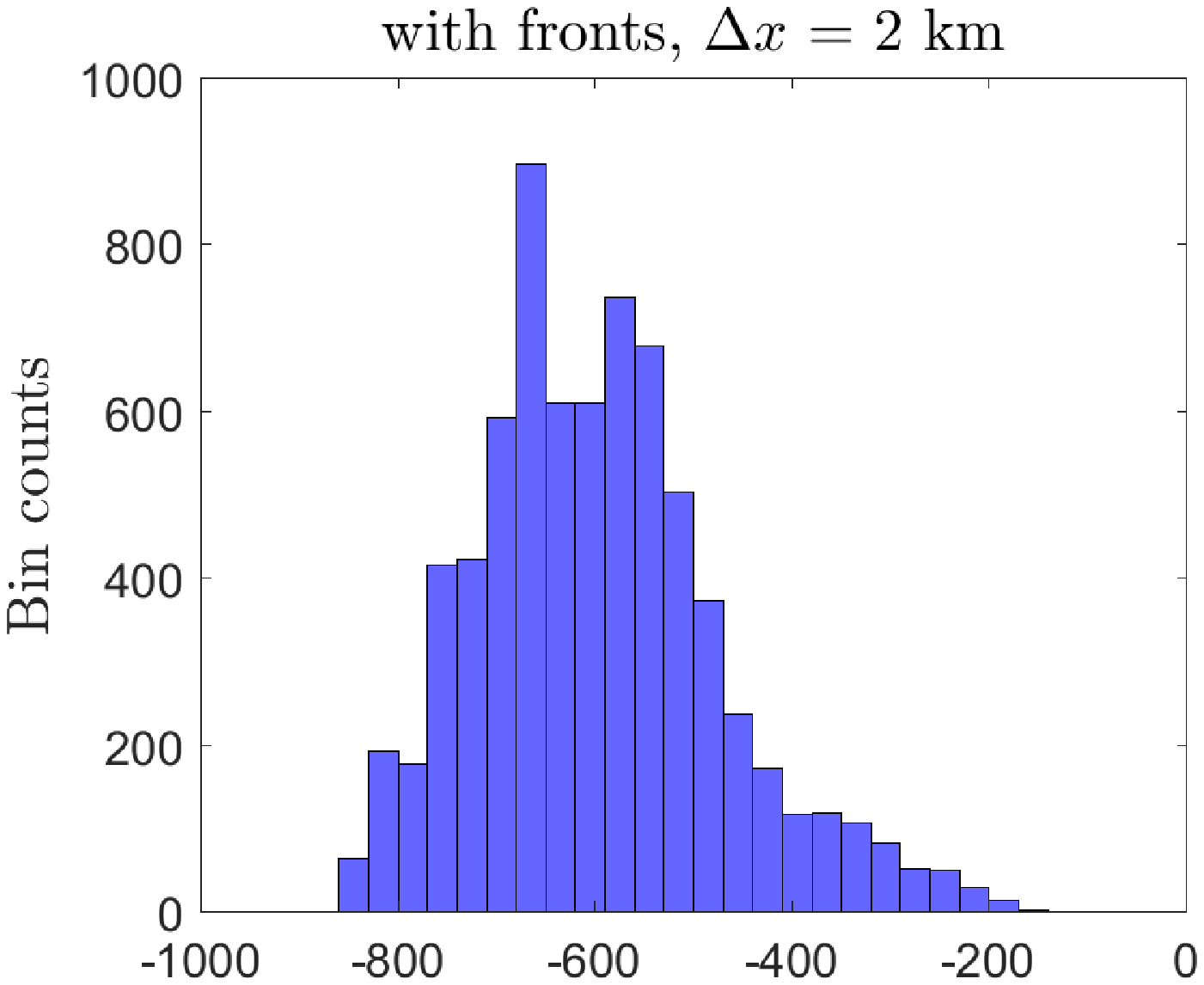}}
		\put(-400,145){(e)} \put(-200,145){(f)} \\
	{\includegraphics[width=2.70in]{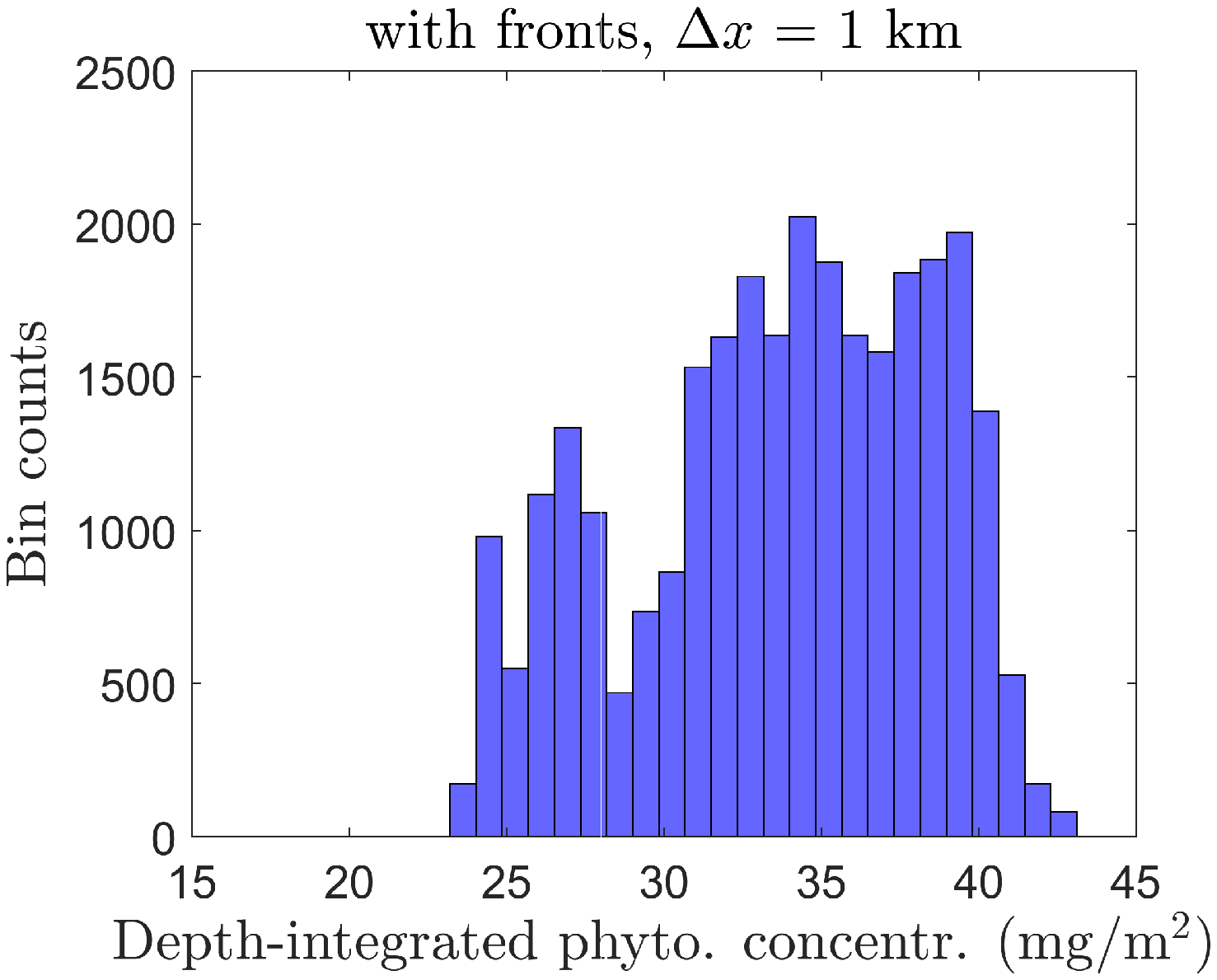}}
	{\includegraphics[width=2.70in]{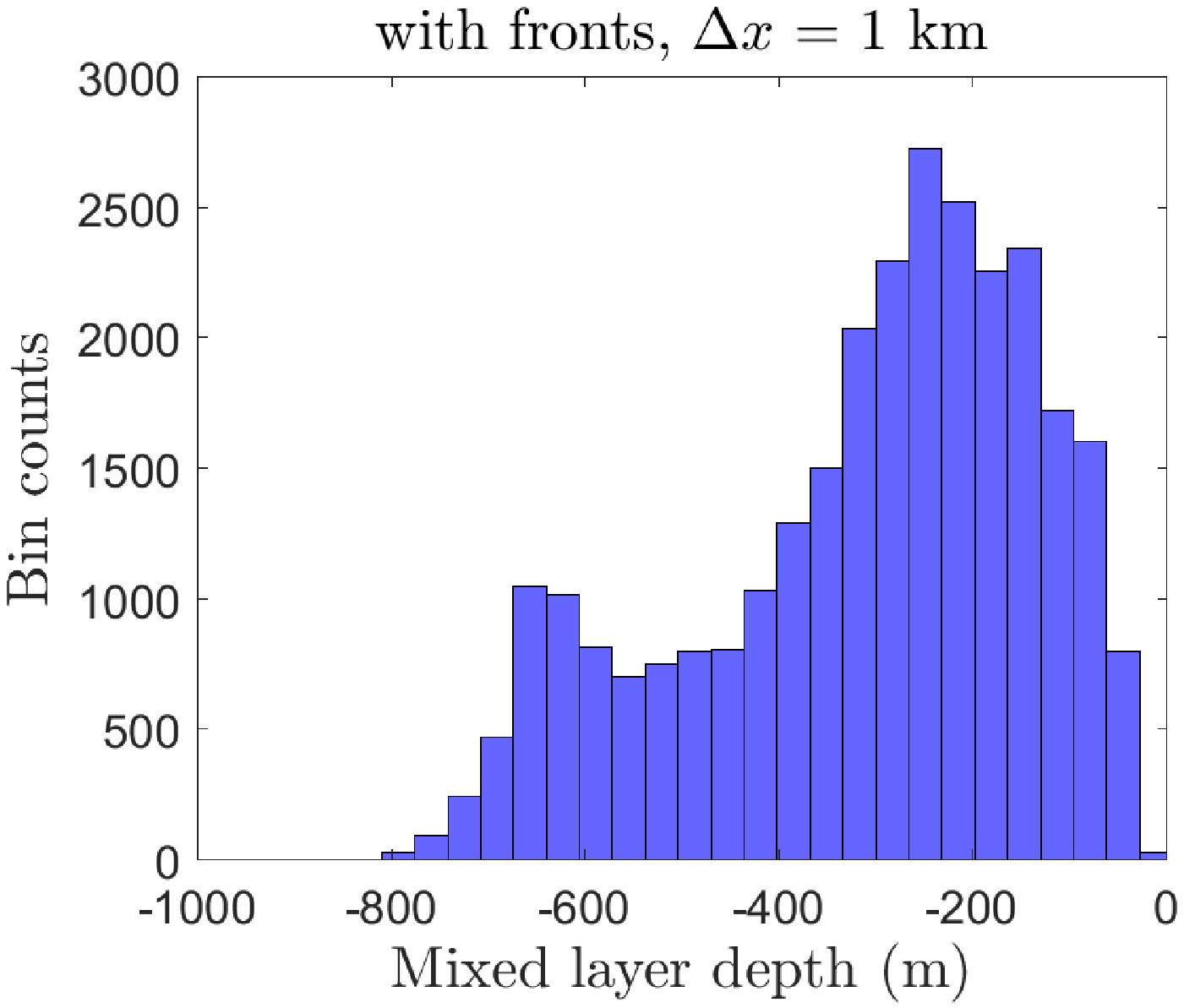}}
		\put(-400,145){(g)} \put(-200,145){(h)} \\
	\caption{ The histogram of  MLD and the phytoplankton concentration integrated down to $z \approx -400~\rm{m}$; for yearday 60  obtained from the numerical simulation for  (a, b) a case without fronts; and different numerical resolutions (c,d) 4 km; (e,f) 2 km; and (g,h) 1 km.}
	\label{fig:hist_phy_res}
\end{figure}

 %
	

Most  models lack the required grid resolution to  resolve submesoscale eddies and frontal processes. We consider four cases for the study of the spatial resolution with $\Delta x =\Delta y=0.75,~1,~2$ and 4 km, where only the 0.75, 1 and 2 km cases have resolutions smaller than  $L_R$. The results are shown in Figure \ref{fig:MLD_phy_num_res}. Also, the case without fronts previously presented in Figure \ref{fig:mld_phy_avg} is added for comparison. Clearly, as shown in Figure \ref{fig:MLD_phy_num_res}a, there is direct relation between the shoaling of the ML and the spatial resolution. The ML is shallower for  numerical simulations with $\Delta x =  0.75$ and 1 km and relatively shallow for 2  km resolution, which can better resolve submesoscale processes. For the coarse  case, the mixed layer is  deep but still shallower than the case without fronts. The effect of numerical resolution and frontal strength on mixed layers  is also reflected in the production of phytoplankton in the winter. In the 0.75 and 1 km resolution simulations, the production of phytoplankton is  more due to the shallower mixed layer, which provides the opportunity for phytoplankton to be exposed to the sunlight for longer time scales and hence grow more. These results highlight the vital role of fronts, their related instabilities and processes on the production of  phytoplankton in the North Atlantic winter and show the importance of proper spatial resolution for  correct prediction of the phytoplankton fate in numerical models.

The standard deviation of the MLD (Figure \ref{fig:MLD_phy_num_res}c) 
shows that the variability of the MLD is higher for higher resolutions, especially there is a noticeable difference between 0.75, 1 and 2 km with 4 km resolutions. The case without fronts has a much smaller standard deviation compared to cases with fronts, highlighting the effect of fronts on the patchiness of the MLD. For phytoplankton concentration, all the cases with fronts basically have similar standard deviation, showing no clear relation between the numerical resolution  and  the variability in  phytoplankton (Figure \ref{fig:MLD_phy_num_res}d). But similar to the MLD, without fronts the standard deviation of phytoplankton concentration is much less than the cases with fronts, signifying the role of fronts on the patchiness of phytoplankton. 
We consider the variability of MLD and phytoplankton for different numerical resolutions by plotting their histograms in Figure \ref{fig:hist_phy_res}. While the increase of numerical resolution leads to higher variability in the MLD, the phytoplankton variability is unchanged. As previously shown in Figure \ref{fig:MLD_phy_num_res}c, phytoplankton variances are similar for different resolutions, however  phytoplankton concentrations from higher resolution models are shifted towards higher values, while phytoplankton concentrations from lower resolution models vary over lower values. This means that the productivity  increases with the increase of spatial resolution. There is also a large difference between the cases with fronts and without fronts for both the MLD and the phytoplankton concentration. The case without fronts shows deep MLD, low phytoplankton concentration and limited variability for both quantities compared to cases with fronts. 

\subsection{Comparison with satellite data}\label{sec:comp_data}
Chlorophyll-a (Chla) data for year 2008 have been obtained from Moderate Resolution Imaging Spectroradiometer (MODIS) aboard NASA's Aqua satellite and  was launched in May 2002. MODIS-Aqua has  two spectroradiometers that operate in two bands from 620 nm to 670 nm and from 841 to 876 nm, each with an along-track and cross-track  resolution of 250 m. These ranges have the sufficient sensitivity for detecting the color changes in the ocean water (\citet{CHEN[2007]}).
 
  \begin{figure}[ht!]	
  	\centering
  	{\includegraphics[width=3.in]{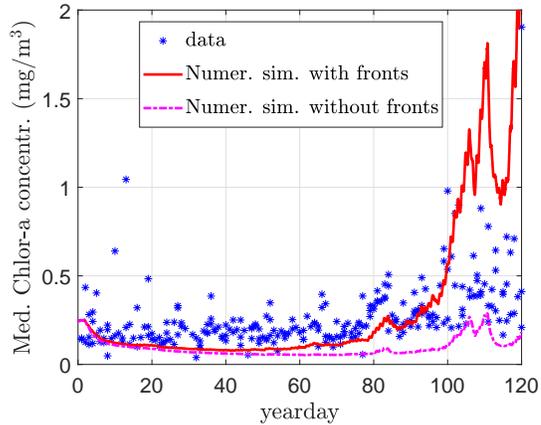}}
  	\caption{ Comparison of the median of Chlorophyll-a concentration  between the longitude $20^{\circ}\rm{W}-45^{\circ}\rm{W}$  and the latitude $47^{\circ}\rm{N}-53^{\circ}\rm{N}$ obtained from MODIS-Aqua  with the median of phytoplankton concentration at $z \approx -5.85 ~\rm{m}$ obtained from the numerical simulation with $\Delta x $ = 1 km. Results for the case without fronts are shown for comparison. Clearly, when fronts are present the simulation results show better agreement with data both in the winter and early spring.}
  	\label{fig:phy_sim_data}
  \end{figure}

Figure \ref{fig:phy_sim_data} compares the time series of the Chl-a median in 2008 from MODIS-Aqua and the median of phytoplankton concentration from the numerical simulation. MODIS-Aqua data are obtained over an area limited between longitude $20^{\circ}\rm{W}-45^{\circ}\rm{W}$ and latitude $47^{\circ}\rm{N}-53^{\circ}\rm{N}$. The  median of the phytoplankton concentration from numerical simulation is calculated at $z \approx -5.85~\rm{m}$, from $x=0-96~\rm{km}$ and $y=100-400~\rm{km}$. Also, phytoplankton concentration from simulations with no fronts is included for comparison.
The figure shows that there is  growth around yearday 77 observed in both the satellite and simulation results. 
Persistent growth occurs after yearday 85 and is in agreement with the increase of $Q_{\rm{net}}$ after around yearday 85, which facilitates the onset of stratification. Although the numerical simulation results show  good agreement with the data, after spring initiation the phytoplankton concentration from simulation is higher than the concentration derived from satellite data. This could be attributed to the absence of grazing of phytoplankton by zooplankton and abundance of nutrients in our model as described in equation (\ref{eq:phy_eq}). Also, we use a constant mortality rate for winter simulations which needs to be modified with the start of spring.

\begin{figure}[ht!]	
 	\centering
 	{\includegraphics[width=2.7in]{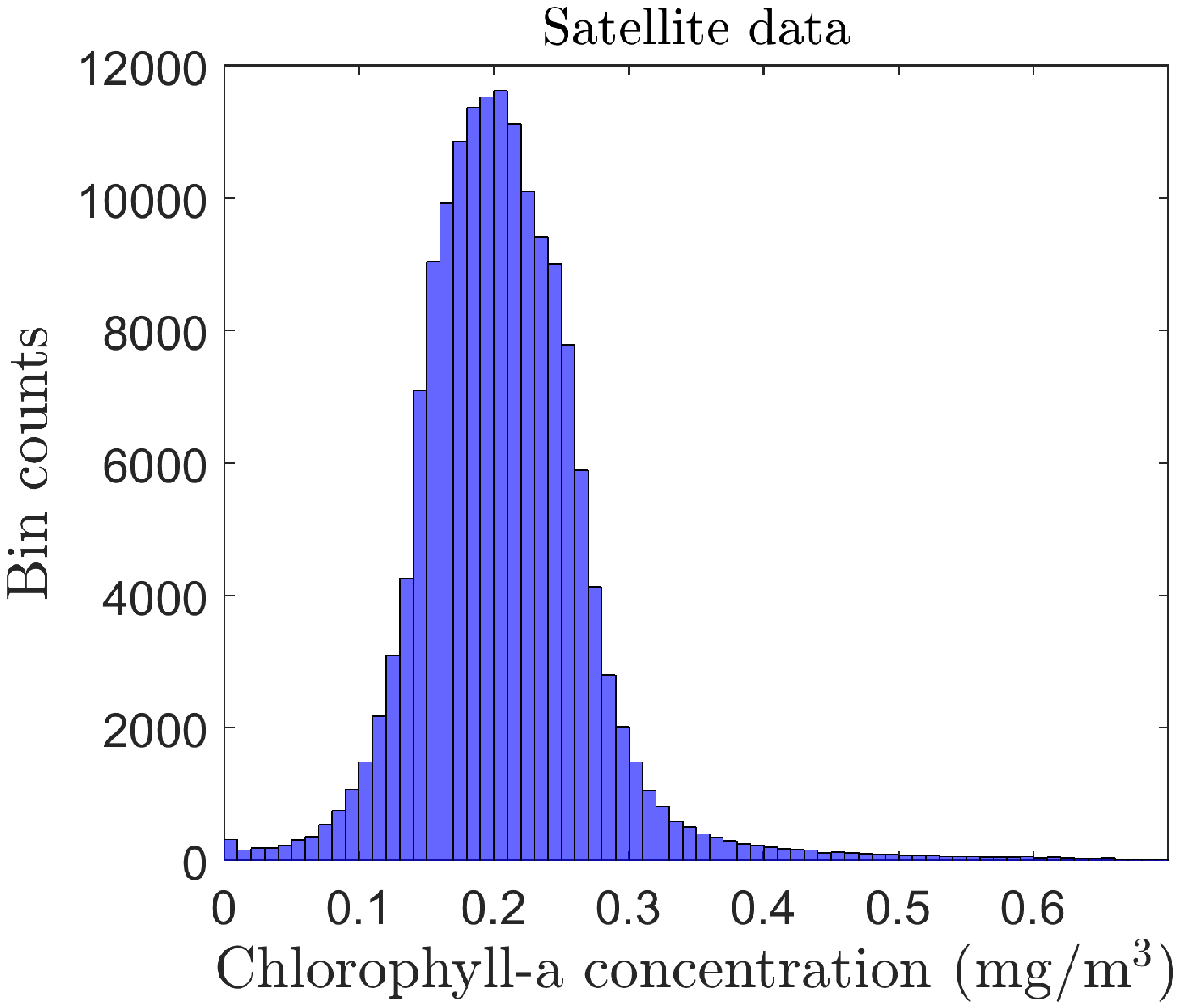}}
 	{\includegraphics[width=2.7in]{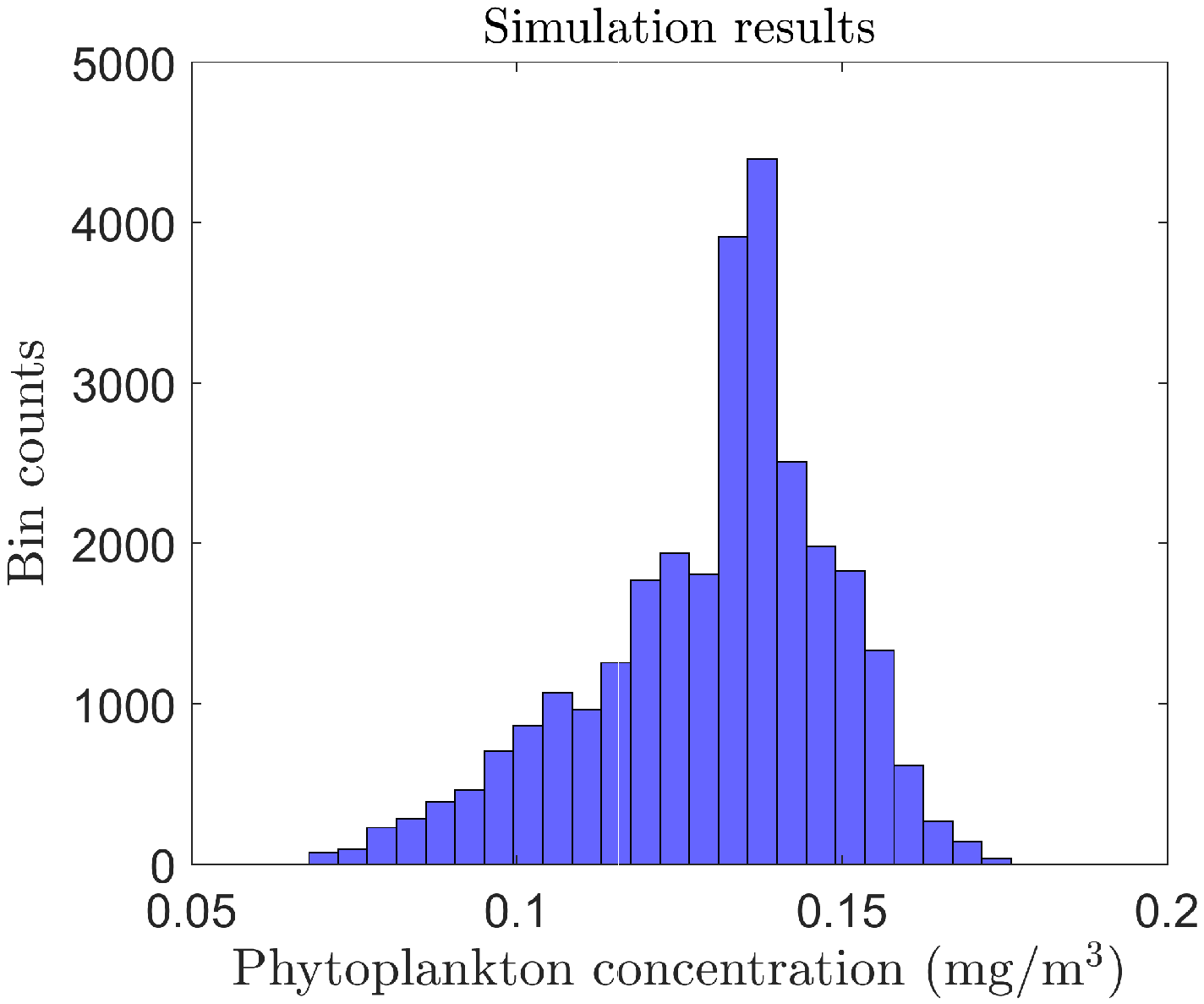}}
 	\put(-360,150){(a)} \put(-170,150){(b)} \\  	
 	\caption{ The histogram of (a) the Chlorophyll-a concentration averaged on longitude $20^{\circ}\rm{W} - 45^{\circ}\rm{W}$  and the latitude $47^{\circ}\rm{N} - 53^{\circ}\rm{N}$ obtained from MODIS-Aqua in 2008; (b) the surface-averaged phytoplankton concentration at $z \approx -5.85~\rm{m}$ obtained from the numerical simulation.}
 	\label{fig:hist_sim_data}
 \end{figure}

In Figure \ref{fig:hist_sim_data}, we compare the histogram of the Chl-a from MODIS-Aqua between longitude   $20^{\circ}\rm{W}-45^{\circ}\rm{W}$ and latitude $47^{\circ}\rm{N}-53^{\circ}\rm{N}$ with the histogram of phytoplankton from the simulation results at $z \approx -5.85~\rm{m}$, from $x= 0-96~\rm{km}$ and $y=100-400~\rm{km}$. The histogram of the satellite Chlorophyll-a concentration in Figure \ref{fig:hist_sim_data}a shows a log-normal distribution that is consistent with \citet{CAMPBELL[1995]}. The simulation results in Figure \ref{fig:hist_sim_data}b look significantly different from data with much smaller variability. 
In our simulations, we solve a simplified model for only one phytoplankton species in a relatively small domain and do not incorporate competition between different types of phytoplankton.
Also, as shown in Figures \ref{fig:MLD_phy_num_res} and \ref{fig:hist_phy_res} the phytoplankton concentration is dependent on numerical resolution. There are many scales between the grid size and phytoplankton scale that our model cannot resolve. Hence, the  grid resolutions we use for the size of our domain cannot capture all the instabilities and turbulent processes that cause the variability in the phytoplankton population.
Also, here we have assumed that zooplankton grazing is negligible,  nutrients are abundant and are uniformly distributed in the mixed layer. All these limitations and simplifications can potentially influence the distribution of  phytoplankton and result in a different variability in the phytoplankton concentration within the numerical simulation domain. 

\section{Concluding remarks}\label{sec:conc}
Our results show that while strong winds and loss of heat in the subpolar North Atlantic during the winter lead to deep mixed layers and keep mixed layer unstratified, the restratifying effect of fronts is at play even in the winter and results in shallower mixed layers, than would be in the absence of fronts.
Further, the simulations show that without fronts the mixed layer deepens constantly during the winter, while with fronts the average mixed layer depth is about 400 m and mostly decreases in the winter. 
The mixed layer depth when fronts are present is even up to about 700 m shallower  than the average mixed layer depth when there are no fronts. 
Fronts also increase patchiness, creating regions of shallow mixed layers across the domain. 
Our simulations show that increase in  frontal strength leads to decrease of the average mixed layer depth (MLD) and increase in the patchiness of the MLD.

Phytoplankton growth substantially increases in winter simulations that include fronts. 
While in many parts of the modeled ocean, phytoplankton dwindle due to deep mixed layers, fronts generate regions of enhanced stratification or shallower mixed layer, where phytoplankton can survive and grow as they have the opportunity to be exposed to the sunlight for longer periods. Moreover, the shallower mixed layer prevents the dilution of phytoplankton and their export to the deep ocean, which is also fundamental for sustaining phytoplankton during the winter. The high concentration and production of phytoplankton near fronts contribute to the sustenance and productivity of phytoplankton during the winter in the North Atlantic, where the light is scant. This sustenance of phytoplankton is essential for providing the seed population for the spring bloom, where the increase of sunlight and heat and hence restratification allow the exponential growth of phytoplankton. Additionally, stronger fronts, which have shallow average MLD, cause higher production of phytoplankton compared to weak fronts. 

Besides fronts, high variability of  air-sea fluxes can cause restratification of the mixed layer and change of the MLD. Simulations reveal that the transient shoaling (or deepening) caused by variable air-sea fluxes has little effect on the production of phytoplankton mainly due to deep mixed layers and short time scales of episodic air-sea fluxes compared to phytoplankton growth time scales, which are on the order of days. Other important reasons for the insensitivity of phytoplankton growth to the variability in  air-sea fluxes are short days and weak light intensity during the winter in the subpolar North Atlantic.

The assumptions used in this idealized study include only one phytoplankton species and constant growth and mortality time scales. Further, zooplankton and nutrients are not explicitly modeled. 
While resolving mixed layer instabilities, our numerical simulations do not capture all small-scale processes that influence mixing and transport in the ocean between the model grid scales and phytoplankton size. It would be interesting in a future study to consider the competition between different phytoplankton species, and a Lagrangian approach to evaluate how fronts and resultant patchiness can increase the residence time of phytoplankton  in regions of shallow mixed layer and contribute to their growth during the winter in the North Atlantic.

\acknowledgments
The simulation data can be obtained from the corresponding author (fkarimpour@umassd.edu). The MERRA-2 data can be obtained from https://gmao.gsfc.nasa.gov/reanalysis/MERRA-2/. The MODIS-Aqua data can be downloaded from https://oceancolor.gsfc.nasa.gov/data/aqua/.
The PSOM code can be downloaded from https://github.com/PSOM/. 
FK and AT acknowledge the support from  the National Science Foundation under grant OCE-1434512. AM acknowledges funding from  the National Science Foundation under grant OCE-1434788.

%
%
%
%
%
%
%
%
%

\end{document}